\documentclass[twocolumn]{aastex6}

\shortauthors{Jiang et al.}
\shorttitle{The Final SDSS $z>5.7$ Quasar Sample}

\usepackage{amsmath}

\newcommand{\lya}{Ly$\alpha$}

\newcommand{\cii}{C\,{\sc ii}}

\newcommand{\civ}{C\,{\sc iv}}
\newcommand{\heii}{He\,{\sc ii}}
\newcommand{\feii}{Fe\,{\sc ii}}

\newcommand{\mgii}{Mg\,{\sc ii}}

\begin{document}

\title{The Final SDSS High-Redshift Quasar Sample of 52 Quasars at $z>5.7$}

\author{Linhua Jiang\altaffilmark{1}, Ian D. McGreer\altaffilmark{2}, 
Xiaohui Fan\altaffilmark{2}, Michael A. Strauss\altaffilmark{3},
Eduardo Ba{\~n}ados\altaffilmark{4,11}, Robert H. Becker\altaffilmark{5},
Fuyan Bian\altaffilmark{6,12}, Kara Farnsworth\altaffilmark{2,5}, 
Yue Shen\altaffilmark{7,8,13}, 
Feige Wang\altaffilmark{1,9}, Ran Wang\altaffilmark{1}, 
Shu Wang\altaffilmark{1,9}, Richard L. White\altaffilmark{10},
Jin Wu\altaffilmark{1,9}, Xue-Bing Wu\altaffilmark{1,9}, 
Jinyi Yang\altaffilmark{1,9}, and Qian Yang\altaffilmark{1,9}}

\altaffiltext{1}{Kavli Institute for Astronomy and Astrophysics, Peking
   University, Beijing 100871, China; jiangKIAA@pku.edu.cn}
\altaffiltext{2}{Steward Observatory, University of Arizona,
   933 North Cherry Avenue, Tucson, AZ 85721, USA}
\altaffiltext{3}{Department of Astrophysical Sciences, Princeton University,
  Princeton, NJ 08544, USA}
\altaffiltext{4}{Observatories of the Carnegie Institution for Science,
	813 Santa Barbara Street, Pasadena, CA 91101, USA}
\altaffiltext{5}{Department of Physics, University of California, Davis, CA 
	95616, USA}
\altaffiltext{6}{Research School of Astronomy and Astrophysics, Australian
   National University, Weston Creek, ACT 2611, Australia}
\altaffiltext{7}{Department of Astronomy, University of Illinois at
   Urbana-Champaign, Urbana, IL 61801, USA}
\altaffiltext{8}{National Center for Supercomputing Applications, University
   of Illinois at Urbana-Champaign, Urbana, IL 61801, USA}
\altaffiltext{9}{Department of Astronomy, School of Physics, Peking
   University, Beijing 100871, China}
\altaffiltext{10}{Space Telescope Science Institute, 3700 San Martin Drive, 
	Baltimore, MD 21218, USA}
\altaffiltext{11}{Carnegie-Princeton Fellow}
\altaffiltext{12}{Stromlo Fellow}
\altaffiltext{13}{Alfred P. Sloan Research Fellow}

\begin{abstract}

We present the discovery of nine quasars at $z\sim6$ identified in the Sloan 
Digital Sky Survey (SDSS) imaging data. This completes our survey of $z\sim6$ 
quasars in the SDSS footprint. Our final sample consists of 52 quasars at 
$5.7<z\le6.4$, including 29 quasars with $z_{\rm AB}\le20$ mag selected from 
11,240 deg$^2$ of the SDSS single-epoch imaging survey (the main survey), 10 
quasars with $20\le z_{\rm AB}\le20.5$ selected from 4223 deg$^2$ of the SDSS 
overlap regions (regions with two or more imaging scans), and 13 quasars down 
to $z_{\rm AB}\approx22$ mag from the 277 deg$^2$ in Stripe 82. 
They span a wide luminosity range of $-29.0\le M_{1450}\le-24.5$. 
This well-defined sample is used to derive the quasar luminosity function 
(QLF) at $z\sim6$. After combining our SDSS sample with two faint 
($M_{1450}\ge-23$ mag) quasars from the literature, we obtain the parameters 
for a double power-law fit to the QLF. The bright-end slope $\beta$ of the QLF 
is well constrained to be $\beta=-2.8\pm0.2$. Due to the small number of 
low-luminosity quasars, the faint-end slope $\alpha$ and the characteristic 
magnitude $M_{1450}^{\ast}$ are less well constrained, with 
$\alpha=-1.90_{-0.44}^{+0.58}$ and $M^{\ast}=-25.2_{-3.8}^{+1.2}$ mag. The 
spatial density of luminous quasars, parametrized as 
$\rho(M_{1450}<-26,z)=\rho(z=6)\,10^{k(z-6)}$, drops rapidly from $z\sim5$ 
to 6, with $k=-0.72\pm0.11$. Based on our fitted QLF and assuming an IGM 
clumping factor of $C=3$, we find that the observed quasar 
population cannot provide enough photons to ionize the $z\sim6$ IGM at 
$\sim90$\% confidence. Quasars may still provide a significant fraction of the 
required photons, although much larger samples of faint quasars are needed for 
more stringent constraints on the quasar contribution to reionization.

\end{abstract}

\keywords
{galaxies: active --- galaxies: high-redshift --- quasars: general 
--- quasars: emission lines}

\section{Introduction}

High-redshift ($z\ge6$) quasars are a powerful tool to study the early 
universe. In recent years, more than 100 quasars at $z>5.7$ have been 
discovered. The Sloan Digital Sky Survey \citep[SDSS;][]{york00} pioneered the 
searches of quasars at these redshifts, followed by the 
Canada-France High-redshift Quasar Survey \citep[CFHQS;][]{wil07}, the UKIRT 
Infrared Deep Sky Survey \citep[UKIDSS;][]{war07}, and the Panoramic Survey 
Telescope \& Rapid Response System 1 \citep[Pan-STARRS1;][]{kai10} 
survey. To date over 40 $z\sim6$ quasars have been discovered based on the 
SDSS imaging data 
\citep[e.g.][]{fan01a,fan03,fan04,fan06a,jiang08,jiang09,jiang15}. The UKIDSS 
has discovered several quasars \citep{ven07,mort09,mort11}, including 
the most distant quasar known at $z=7.08$ \citep{mort11,bar15}. 
The CFHQS found 20 
quasars over $\sim500$ deg$^2$ of sky \citep{wil07,wil09,wil10}. 
The Pan-STARRS1 covers 3$\pi$ steradians of the sky, and is now producing a
large number of high-redshift quasars \citep{morg12,ban14,ven15,ban16},
including three quasars at $6.5<z<6.7$ \citep{ven15}.
Most recently, the VISTA Kilo-Degree Infrared Galaxy (VIKING) survey, 
the Dark Energy Survey (DES), the VST ATLAS survey, and the Subaru High-$z$ 
Exploration of Low-Luminosity Quasars (SHELLQ) project, have started to yield 
$z\ge6$ quasars \citep[e.g.][]{ven13,car15,reed15,mat16}. 
The number of high-redshift quasar discoveries is increasing steadily.

Meanwhile, the sample of bright high-redshift quasars, especially luminous 
SDSS quasars, have been studied extensively in multiple wavelength bands from 
X-ray to radio. These quasars are very luminous with $M_{1450}<-26$ mag. 
Deep optical spectra have revealed strong or even complete absorption in the 
Ly$\alpha$ forests, indicating that the redshift probed ($z\sim6$) is close to 
the epoch of cosmic reionization 
\citep[e.g.][]{bec01,whi03,fan06b,car10,bol11,mcg15}. 
Their infrared spectroscopy show that these 
luminous quasars harbor billion-solar-mass black holes and emit near the
Eddington limit, suggesting the rapid growth of central black holes at this
early epoch \citep[e.g.][]{jiang07,kurk07,wil10a,derosa14,jun15,wu15}. 
The broad emission lines of these quasars exhibit solar or supersolar 
metallicity, implying that vigorous star formation and element enrichment have 
occurred in their broad-line regions \citep[e.g.][]{jiang07,jua09,derosa11}. 
In addition, observations in the mid/far-IR, mm/sub-mm, and radio wavebands
have provided rich information about the dust emission and star formation
in the host galaxies \citep[e.g.][]{jiang06,jiang10,wal09,gal10,wang11,wang13,car13,omo13,cal14,lei14,ban15,lyu16}.
Therefore, high-redshift quasars are a powerful probe for understanding black 
hole accretion, galaxy evolution, and the intergalactic medium (IGM) state in 
the first billion years of cosmic time.

In this paper, we present nine quasars newly found in the SDSS, including 
seven quasars in the SDSS main survey area, one quasar in the SDSS overlap 
regions, and one quasar in SDSS Stripe 82. The overlap regions are the regions 
with overlapping imaging in the SDSS, which results in multiple observations 
of individual sources within these regions. Stripe 82 covers $\sim300$ 
deg$^2$, and was repeatedly scanned 70--90 times by the SDSS imaging survey. 
We describe these regions in Section 2. With the discovery of these nine 
quasars, we have completed our survey of $z\sim6$ quasars in the SDSS 
footprint. We summarize our survey of SDSS quasars in the second half of the 
paper. With a total of 52 quasars, we derive the quasar luminosity function 
(QLF) at $z\sim6$, and in particular, improve the measurement of the QLF at 
the bright end.

The layout of the paper is as follows. In Section 2, we review our survey of
$z>5.7$ quasars in the SDSS. In Section 3, we present the nine new quasars. 
In Section 4, we summarize our complete sample of 52 SDSS quasars and 
calculate the QLF at $z\sim6$. In Section 5, we discuss the evolution of 
luminous quasars at high redshift and the quasar contribution to cosmic
reionization at $z\sim6$. We summarize the paper in Section 6. 
Throughout the paper, SDSS magnitudes are expressed in the AB system. Near-IR 
and mid-IR magnitudes are in the Vega system. We use a $\Lambda$-dominated
flat cosmology with $H_0=70$ km s$^{-1}$ Mpc$^{-1}$, $\Omega_{m}=0.3$, and
$\Omega_{\Lambda}=0.7$.

\section{Survey of $z>5.7$ quasars in the SDSS}

In this section, we briefly review our survey of $z>5.7$ quasars selected in 
the SDSS. We will need this information for Sections 3 and 4.
The SDSS is an imaging and spectroscopic survey of the sky using
a dedicated wide-field 2.5 m telescope \citep{gun06} at Apache Point
Observatory. Imaging was carried out in drift-scan mode using a 142 mega-pixel
camera \citep{gun98} which gathered data in five broad bands, $ugriz$, 
spanning the range from 3000 to 10,000 \AA\ \citep{fuk96}, on moonless 
photometric \citep{hog01} nights of good seeing. The effective exposure time 
was 54.1 seconds. An SDSS run (strip) consists of 6 parallel scanlines (camera 
columns) for each of the five $ugriz$ bands. The scanlines are $13\farcm5$
wide with gaps of roughly the same width, so two interleaving strips make a
stripe. SDSS scanlines are divided into fields, and a field is the union of
five $ugriz$ frames covering the same region of sky.
The images were processed using specialized software \citep{lup01},
and are photometrically \citep{tuc06,ive04,pad08} and astrometrically
\citep{pier03} calibrated using observations of a set of primary standard 
stars \citep{smi02} on a neighboring 20-inch telescope.

\subsection{Quasars in the SDSS Main Survey}

The initial goal of the SDSS imaging survey was to scan 8500 deg$^2$ of the 
north Galactic cap. The total unique area was expanded to 14,555 deg$^2$, by 
adding $>$5000 deg$^2$ in the south Galactic cap (SGC) \citep{aih11}. 
\citet{fan01a,fan03,fan04,fan06a} discovered 19 $z\sim6$ quasars from the SDSS
photometry, primarily in the north Galactic cap. Most of these quasars are 
bright ($z_{\rm AB}\le20$ mag), and were selected from single-epoch imaging 
data (hereafter referred to as the SDSS main survey). They represent the most 
luminous quasars at $z\ge6$. However, there were main survey regions remaining 
unsearched, particularly in the SGC. In this paper we report on the discovery 
of additional quasars found in these regions. 

\begin{figure} 
\epsscale{1.2}
\plotone{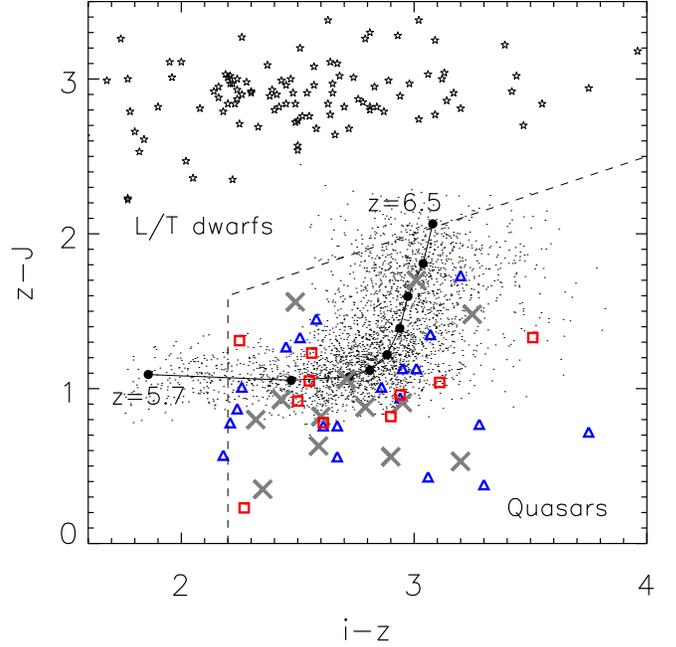}
\caption{The $z_{AB}-J$ versus $i_{AB}-z_{AB}$ color-color diagram for quasar 
candidate selection. The open stars represent a sample of known L/T dwarfs 
drawn from DwarfArchives.org. The black dots represent simulated quasars 
(Section 4.3) with a luminosity of $M_{1450}\approx-26$ mag at $5.7<z<6.5$. 
No photometric errors are added. The black circles show the median track of 
quasar colors. The dashed lines indicate our selection criteria. The blue
triangles, gray crosses, and red squares represent the SDSS quasars that have 
$J$-band photometry in the main survey, Stripe 82, and overlap regions, 
respectively.
\label{fig:colorselect}}
\end{figure}

The quasar selection procedure in the main survey has been discussed in detail 
in the papers mentioned above. Here we briefly review the procedure. 
Because of the rarity of high-redshift quasars and overwhelming number of
contaminants, the procedure consists of four basic steps. The first step is to 
select $i$-band dropout objects mainly in high galactic latitude $|b|>30$. 
Sources with $i_{AB}-z_{AB}>2.2$ mag and $z$-band error $\sigma_z<0.1$ mag 
(roughly $z_{AB}\le20$ mag) that were not detected in the $ugr$ bands are 
selected as $i$-dropout objects. The simple color cut $i_{AB}-z_{AB}>2.2$ is 
used to separate quasars (and cool brown dwarfs) from the majority of stellar 
objects \citep[e.g.][]{fan99,str99}. Beyond the limit of $\sigma_z<0.1$ mag, 
the number of contaminants increases dramatically. The second step is to 
remove false $i$-dropout objects and improve photometry. All $i$-dropout 
objects are visually inspected, and false detections such as cosmic rays are
removed. If necessary, we also take deeper imaging data to improve the $i$ and 
$z$-band photometry to reduce the number of contaminants. The third step is to 
take near-IR (usually $J$ band) photometry of $i$-dropout objects with another
telescope. In the $z_{AB}-J$ versus $i_{AB}-z_{AB}$ color-color diagram, 
high-redshift quasar candidates are separated from brown dwarfs. Specifically, 
quasar candidates satisfy the criterion $z_{AB}-J < 0.5\,(i_{AB}-z_{AB})+0.5$
(see also Figure \ref{fig:colorselect}). The final step is to take 
spectroscopic observations and identify quasar candidates.

In addition to the above `standard' survey to a limit of $\sim10\sigma$ 
detections in the SDSS $z$-band images, we also selected two small samples of 
quasar candidates using a `non-standard' method. The first sample consisted of 
candidates down to $\sim7\sigma$ in the $z$-band images in part of the UKIDSS 
footprint. This is similar to the test done by \citet{fan06a}. 
We used a more stringent color cut 
$i_{AB}-z_{AB}>2.5$ mag to reduce the number of contaminants caused by larger 
photometric uncertainties. We further required that the candidates should be
detected at a significance level of $>7\sigma$ in the UKIDSS $Y$ and $J$ 
bands. Two of the quasars in this paper were selected using this method.
The second `non-standard' sample consisted of several candidates 
with $i_{AB}-z_{AB}$ colors between 2.1 and 2.2 mag, slightly bluer than that
used for the `standard' survey. 
One quasar in this paper was selected using this method.

\subsection{Quasars in the SDSS Stripe 82}

In addition to the single-epoch main imaging survey, the SDSS also conducted a 
deep survey by repeatedly imaging a $\sim300$ deg$^2$ area on the Celestial 
Equator in the south Galactic cap \citep{ade07,ann14,jiang14}. 
This deep survey stripe, or Stripe 82, roughly spans $\rm 20^h<R.A.<4^h$ and 
$\rm -1.26\degr<Decl.<1.26\degr$, and was scanned 70--90 times in total, 
depending on R.A. along the stripe. The multi-epoch data have been used to
construct co-added images that can reach two magnitudes deeper than
the SDSS single-epoch images \citep[e.g.][]{ann14,jiang14,fli16}. 
Using these co-added data, we have found 12 $z>5.7$ quasars in Stripe 82 
\citep{jiang08,jiang09}. These quasars have $20<z_{\rm AB}<22$ mag, and are on 
average two magnitudes fainter than those found in the main survey. 
In this paper we present one new quasar found in Stripe 82.
The quasar selection procedure for Stripe 82 is very similar to that for 
the main survey, except that the survey limit is $z_{AB}\sim22$ mag 
($\sigma_z<0.11$ mag) instead of $z_{AB}\sim20$ mag. 

\subsection{Quasars in the SDSS Overlap Regions}

We have also carried out searches of $z\sim6$ quasars in the SDSS overlap 
regions, the regions that were scanned by two or more SDSS imaging runs.
The SDSS imaging runs generally overlap each other, due to the survey geometry 
and strategy. The imaging survey in drift-scan mode was along great circles, 
and had two common poles. The fields overlap more substantially when they 
approach the survey poles. In addition, the two interleaving strips that make 
any stripe overlap slightly, leading to repeat observations in a small area. 
Furthermore, if the quality of a run, or part of a run, did not meet the SDSS 
standard seeing and photometric criteria, the relevant region
was re-observed, yielding duplicate observations in this region. 
The total area of the overlap regions is more than one-fourth of the SDSS 
footprint. 
These overlap regions provide a unique dataset that allows us to select 
high-redshift quasars more than 0.5 mag fainter (in the $z$ band) than those 
found with the SDSS single-epoch data. 
We have discovered eight quasars in the overlap regions \citep{jiang15}.

The selection procedure of overlap-region quasars is slightly different from 
those described above. The image quality is usually different between the 
repeat runs. In the first step of the selection procedure, the magnitude limit 
for both `primary' and `secondary' detections is $z_{AB}<20.7$ mag or 
$\sigma_z<0.155$ mag ($7\sigma$ detection). The candidates are actually 
fainter than $10\sigma$ detections, because otherwise they would have been 
selected in the main survey. We focused on high galactic 
latitude ($|b|>30$) regions. Using repeat observations ensures that most  
$i$-dropout objects we select are physical sources, rather than artifacts or
cosmic rays. In the second step, we take deeper $i$ and $z$ band images to 
improve photometry for $i$-dropout objects. The rest of the selection
procedure remains the same. The details are given in \citet{jiang15}.

\section{Discovery of nine new quasars}

In this section we present the discovery of nine new quasars in the SDSS.
The basic information of the quasars, including their coordinates, redshifts,
and broad-band ($izJ$) photometry, is given in Table 1. One of them (SDSS 
J083525.76+321752.6; hereafter we use J0835+3217 for brevity) is found in 
the overlap regions, and another one SDSS J211951.89--004020.1 (hereafter
J2119--0040) is found in Stripe 82. The other seven quasars were found based 
on the SDSS single-epoch data. The naming convention for SDSS sources is
SDSS JHHMMSS.SS$\pm$DDMMSS.S, and the positions are expressed in J2000.0
coordinates. For brevity, we use JHHMM$\pm$DDMM in the following text.

\floattable
\begin{deluxetable}{cCCCCl}
\tablecaption{Nine New Quasars in the SDSS}
\tablewidth{0pt}
\tablehead{\colhead{Quasar (SDSS)} & \colhead{Redshift} & 
	\colhead{$i_{\rm AB}$} & \colhead{$z_{\rm AB}$} & 
	\colhead{$J_{\rm Vega}$} & \colhead{Notes}}
\colnumbers
\startdata
J081054.32+510540.1 & 5.80\pm0.03 & 21.52\pm0.13 & 19.34\pm0.07 & 18.77\pm0.06 & See also \citet{ban16} \\
J083525.76+321752.6 & 5.89\pm0.03 & >23.0        & 20.73\pm0.20 & 20.50\pm0.20 & Overlap regions \\
J114338.34+380828.7 & 5.81\pm0.03 & 21.97\pm0.19 & 19.76\pm0.09 & 18.98\pm0.09 & See also \citet{ban16} \\
J114803.28+070208.3 &6.339\pm0.001& 23.20\pm0.35 & 20.79\pm0.10 & 19.36\pm0.11 & See also S. Warren et al. (in prep.) \\
J124340.81+252923.9 & 5.85\pm0.03 & 23.08\pm0.29 & 20.22\pm0.10 & 19.21\pm0.12 & See also S. Warren et al. (in prep.) \\
J160937.27+304147.7 & 6.16\pm0.03 & >22.5        & 20.26\pm0.13 & 19.39\pm0.14 & See also S. Warren et al. (in prep.) \\
J162100.92+515548.8 & 5.71\pm0.03 & 21.86\pm0.13 & 19.70\pm0.07 & 19.11\pm0.20 &   \\
J211951.89$-$004020.1& 5.87\pm0.03 & 23.99\pm0.27 & 21.68\pm0.10 & 20.87\pm0.12 & Stripe 82 \\
J231038.88+185519.7 & 6.003\pm0.001& 21.66\pm0.25 & 19.21\pm0.09 & 17.94\pm0.05 &  \\
\enddata
\end{deluxetable}

\subsection{Observations and Data Reduction}

We first present the observations of quasar candidates in Stripe 82, which
were done in 2009 and 2010. The $J$-band photometry of $i$-dropout objects
(quasar selection procedure step 3) was made using the SAO Widefield InfraRed 
Camera \citep[SWIRC;][]{swirc} on the MMT. The observing strategy is the
same as that of \citet{jiang08,jiang09}. The observing conditions were typical,
with relatively clear skies and $\sim 1\farcs0$ seeing.
The images were reduced using standard IRAF\footnote{IRAF
is distributed by the National Optical Astronomy Observatory, which is
operated by the Association of Universities for Research in Astronomy (AURA)
under cooperative agreement with the National Science Foundation.} routines.
We used bright UKIDSS or 2MASS \citep{skr06} point sources in the same images
for flux calibration. Based on the $J$-band photometry, the final sample of
quasar candidates was selected. We then used the MMT Red Channel 
Spectrograph \citep[RCS;][]{mmtred} to identify these
candidates. The exposure time for each target was 20--30 minutes, depending
on the object brightness and weather conditions. If a target was identified as
a quasar, several further exposures were taken to improve the spectral
quality. The MMT RCS data were reduced using standard IRAF routines.

The observations of quasar candidates in the SDSS overlap regions were
conducted in 2015 and 2016. Deeper $i$ and $z$-band photometry of
$i$-dropout objects (quasar selection procedure step 2) was made using the
wide-field optical imager 90Prime on the 2.3m Bok telescope. The 90Prime
images were reduced in a standard fashion using our own {\tt IDL} routines.
The details of the Bok observations and data reduction can be found in
\citet{jiang15}.
For the $J$-band photometry, we used the UKIDSS data for any candidates that
have significant ($>7\sigma$ in $J$) detections in the UKIDSS. For the other
candidates, we obtained their $J$-band photometry using the MMT SWIRC.
We used the MMT RCS and the Double Spectrograph (DBSP) on the Hale 5.1m
telescope to identify quasar candidates and obtain high-quality optical
spectra. The Hale DBSP data were reduced using standard IRAF routines as well.

The observations of quasar candidates in the SDSS main survey were done
between 2010 and 2015, except for J1621+5155, which was observed in 2006.
We used the Bok/90Prime to improve the $i$ and $z$-band
photometry for the sample of $i$-dropout objects with $\sigma_z>0.1$ mag. 
For the $i$-dropout objects with $\sigma_z<0.1$ mag, we simply used the SDSS
data. The $J$-band photometry was made using the MMT SWIRC, or from the 
UKIDSS archive for the objects detected at $>7\sigma$ in the $J$ band.
We then used the MMT RCS and the Hale DBSP to identify quasar candidates and 
obtain high-quality optical spectra, as we did for the candidates in the 
overlap regions. In addition, we took a deep optical spectrum for 
J2310+1855 in long slit mode using the 
Multi-Object Double Spectrograph (MODS) on the Large Binocular Telescope 
(LBT). The MODS spectra were reduced using standard IRAF routines.

We also obtained deep near-IR spectra for two quasars, J2310+1855 and 
J1148+0702, using Gemini GNIRS and Magellan FIRE, respectively. The GNIRS 
observation of J2310+1855 is part of our large Gemini GNIRS campaign of 
$\sim60$ quasars (GN-2015B-LP-7). The GNIRS campaign is used to measure
the rest-frame properties of a large sample of $z\sim6$ quasars, including
UV continuum slopes, broad emission line properties, black hole masses and 
mass function, etc. Both GNIRS and FIRE (in echelle mode)
provide a simultaneous wavelength coverage of 
$0.9-2.5\mu$m in cross-dispersion mode. The observing strategies for the two
observations were the same. We used the standard ABBA nodding sequence 
between exposures. The exposure time at each nod position was 5 minutes, and 
the distance between the two positions was $2\arcsec$. Before or after the 
exposure of each quasar, a nearby A or F spectroscopic standard star was 
observed for flux calibration and to remove telluric atmosphere absorption.
The GNIRS spectra were reduced using the IRAF Gemini package, and the details
can be found in \citet{jiang07}. The FIRE spectra were reduced using an
{\tt IDL} pipeline developed by the FIRE instrument team, and the details
can be found in \citet{sim11}.

\subsection{Results}

From the above observations, we took spectra for about 30 candidates, and
identified nine quasars, including one quasar (J0835+3217) in the SDSS overlap 
regions, one quasar (J2119--0040) in Stripe 82, and seven quasars in the main 
survey. Table 1 lists the coordinates, redshifts, and the broad-band 
photometry of the quasars. Column 1 shows the J2000 
coordinates, or the source names. Column 2 shows the redshifts, which span
the range $5.7<z<6.4$. The redshifts were mostly measured from the \lya\ 
emission lines, or from the wavelength where the sharp flux decline occurs.
The measurements can be slightly biased towards higher redshifts due to the
\lya\ forest. The redshift error of 0.03 quoted in Column 2 is simply the 
scatter in the relation between \lya\ redshifts and systemic redshifts at low 
redshift \citep[e.g.][]{shen07}. The uncertainties from our fitting process 
and wavelength calibration are negligible in comparison. The redshift of 
J1148+0702 is measured from its \mgii\ emission line (see Section 3.2.1).
The redshift of J2310+1855 is measured from the CO (6--5) observations by 
\citet{wang13}. Columns 3 through 5 show the $i$, $z$, and 
$J$-band photometry. The $i$ and $z$-band photometry was taken from the SDSS, 
or improved by the Bok 90Prime. The $J$-band photometry was taken from the 
UKIDSS, or obtained from the MMT SWIRC. These quasars span a brightness range 
of $19.21<z_{\rm AB}<21.68$ and a luminosity range of $-27.61<M_{1450}<-24.73$ 
mag. 

Among the seven quasars found in the main survey area, three quasars were
selected using the `non-standard' method mentioned in Section 2.1. J1148+0702 
and J1609+3041 are fainter than a $10\sigma$ detection in the SDSS $z$-band 
images, and J1621+5155 has an $i_{AB}-z_{AB}=2.16$ color slightly bluer than 
the 2.2 mag limit. Two quasars, J0810+5105 and J1143+3808, in Table 1, were 
independently discovered by the Pan-STARRS1 \citep{ban16}, as indicated in the 
last column. In addition, J1148+0702, J1243+2529, and J1609+3041 were 
independently discovered by the UKIDSS (S. Warren et al., in preparation; 
see also \citet{mort15}). We also recovered J0100+2802 at $z=6.30$ discovered 
by \citet{wu15}, and two quasars, J1545+6028 at $z=5.78$ and J2325+2628 at 
$z=5.77$, found by \citet{wang16a}. We missed J2356--0622 at $z=6.15$ in
\citet{wang16a}, because this quasar has $\sigma_z=0.12$ mag in the SDSS.

\begin{figure}
\epsscale{1.2}
\plotone{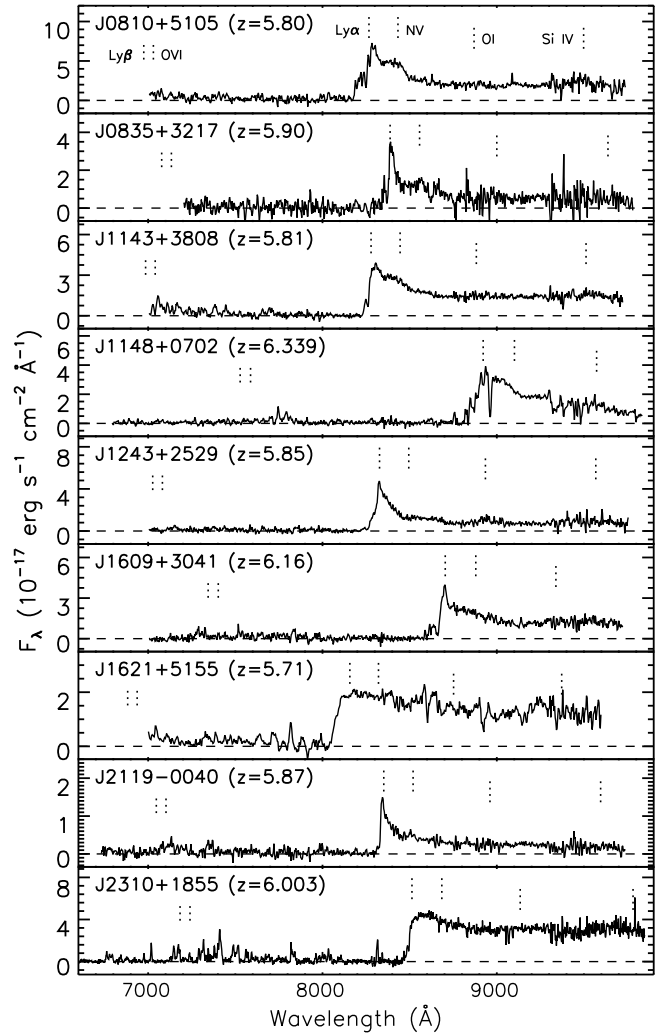}
\caption{Optical spectra of the nine newly discovered quasars. The spectrum of 
J2310+1855 was taken with the LBT MODS. The other spectra were taken with the 
MMT RCS. The dashed lines indicate the zero flux level for each spectrum. 
Each spectrum has been scaled to match the corresponding $z$-band magnitude in 
Table 1, thereby placing it on an absolute flux scale.
\label{fig:spec}}
\end{figure}

Figure \ref{fig:spec} shows the optical spectra of the nine quasars. 
All spectra except J2310+1855 were observed with the MMT RCS. The total 
integration time per object except J2119--0040 was from 40 min to 80 min (20 
min exposures), depending on the quasar brightness and observing conditions. 
The total integration time for the faintest quasar J2119--0040 was 150 min, 
composed of five 30 min exposures. The spectrum of J2310+1855 was obtained 
from the LBT MODS, and the total integration time was 60 min. Each spectrum in 
Figure \ref{fig:spec} has been scaled to match the corresponding $z$-band 
magnitude in Table 1, thereby roughly placing it on an absolute flux scale 
(although variability introduces uncertainty into this calibration).

\begin{figure}
\epsscale{1.2}
\plotone{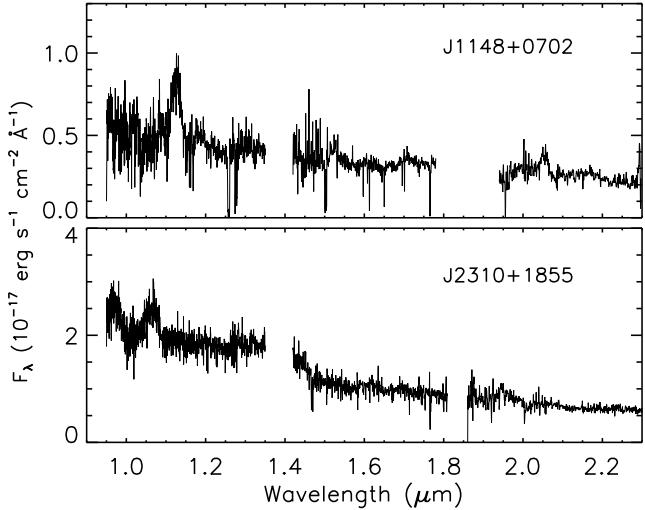}
\caption{Near-IR spectra of J1148+0702 and J2310+1855. The spectrum of 
J1148+0702 was taken with Magellan FIRE, and the total on-source integration 
time was 60 min (12 five-min exposures). The spectrum of J2310+1855 was 
obtained from Gemini/GNIRS, and the integration time was also 60 min. 
Both spectra have been scaled to match the corresponding $J$-band magnitude 
in Table 1.
\label{fig:nirspec}}
\end{figure}

The quasar rest-frame UV spectrum, from the \lya\ emission line to the \feii\ 
bump at $2000\sim3000$ \AA, contains strong diagnostic emission lines and 
provides key information on the physical conditions and emission mechanisms of 
the broad-line region. The rest-frame UV band is
redshifted to the near-IR range for $z\ge6$ quasars. 
As we mentioned earlier, we also obtained near-IR spectra for J1148+0702 
(the highest-redshift quasar in our sample) and J2310+1855 (the most luminous
quasar of the nine) using Magellan FIRE and Gemini GNIRS, respectively.
Figure \ref{fig:nirspec} shows the two near-IR spectra. The total on-source 
integration time for each object was 60 min, broken into 12 five-minute 
exposures. Each spectrum has been scaled to match the corresponding 
$J$-band magnitude in Table 1.

\subsubsection{Notes on individual objects}

{\it J0810+5105 (z=5.80), J1143+3808 (z=5.81), and J1243+2529 (z=5.85)}. 
These quasars were discovered in the SDSS main survey. They are 
at relatively low redshift $\sim5.8$. They all have prominent \lya\ emission.
J0810+5105 has $z_{\rm AB}=19.34$ mag, making it one of
the brightest $z\sim6$ quasars known.

{\it J0835+3217 (z=5.89)}.
J0835+3217 was discovered in the SDSS overlap regions. It is relatively faint 
($z_{\rm AB}=20.73$ mag) compared to those found in the main survey. 
J0835+3217 has a narrow \lya\ emission line. \citet{jiang15} reported on the 
discovery of eight quasars in the overlap regions, and also recovered eight 
previously known quasars in the same area. J0835+3217 is the last one that we 
found in the overlap regions.

{\it J1148+0702 (z=6.339) and J1609+3041 (z=6.16)}.
J1148+0702 and J1609+3041 were selected in a `non-standard' way, as mentioned 
in Section 2.1. They are fainter than $10\sigma$ detections in the SDSS 
$z$-band images. J1148+0702 and J1609+3041 are the two highest-redshift 
quasars in this sample. J1148+0702 is the second highest-redshift quasar found 
in the SDSS. Using the near-IR spectrum in Figure \ref{fig:nirspec}, we 
estimate its central black hole mass based on the empirical scaling relations 
\citep{shen12}. The masses from \mgii\ and \civ\ are 
$(1.26\pm0.14) \times 10^9$ M$_\sun$ and $(2.04\pm0.11) \times 10^9$ M$_\sun$, 
respectively. The redshift estimated from \mgii\ is $6.339\pm0.001$.

{\it J2119--0040 (z=5.87)}.
J2119--0040 is the faintest quasar in our sample, found in the SDSS Stripe 82.
It has strong \lya\ emission.
We previously discovered 12 quasars in Stripe 82 \citep{jiang08,jiang09}.
J2119--0040 is the last one that we found in this area.
The quasars in Stripe 82 form a statistically complete sample down to
$z_{\rm AB}\sim22$ mag.

{\it J1621+5155 (z=5.71)}.
J1621+5155 was selected in a `non-standard' way, with $i_{AB}-z_{AB}<2.2$ mag.
It thus has the lowest redshift in our sample. It is a weak line quasar 
without obvious \lya\ emission in Figure \ref{fig:spec}.
It was not detected in moderate deep millimeter and radio observations
\citep{wang08}.

{\it J2310+1855 (z=6.003)}.
J2310+1855 is the brightest quasar in our sample. It is also one of the most
luminous quasars in the full SDSS $z\sim6$ quasar sample. 
It has very weak \lya\ emission. Weak line quasars seem to be common at 
$z\sim6$ \citep[e.g.][]{ban14,jiang15}.
This quasar has been studied extensively in the
mm/submm and radio bands \citep[e.g.][]{wang13}. Its strong detections of the
[\cii] 158 $\mu$m and CO (6--5) lines provide a redshift measurement of 
$z=6.003\pm0.001$. The redshift calculated from it \mgii\ emission line is 
$z=5.962\pm0.007$. We estimate its central black hole mass from its near-IR
spectrum shown in Figure \ref{fig:nirspec}, and the masses from \mgii\ and 
\civ\ are $(4.17\pm 1.02) \times 10^9$ M$_\sun$ and 
$(3.92\pm 0.48) \times 10^9$ M$_\sun$, respectively.

\clearpage
\floattable
\begin{deluxetable*}{cccCCCCCCCCCcc}
\tabletypesize{\scriptsize}	
\tablecaption{The Final Sample of 52 SDSS Quasars at $z>5.7$}
\tablewidth{0pt}
\tablehead{\colhead{No.} & \colhead{Quasar (SDSS)} & \colhead{Redshift} &
   \colhead{$i_{\rm AB}$\tablenotemark{a}} & \colhead{$z_{\rm AB}$} & 
	\colhead{$Y$} & \colhead{$J$} & \colhead{$H$} & \colhead{$K$} & 
	\colhead{$W1$} & \colhead{$W2$} & \colhead{$M_{1450}$} & 
	\colhead{Discovery paper} & \colhead{Region\tablenotemark{b}}}
\colnumbers
\startdata
1  &  J000239.40+255034.8 & 5.82  & 21.56 & 18.99 & \ldots& >16.5 & \ldots& \ldots& 16.20 & 15.45 & -27.61 &   \citet{fan04} &  M+O  \\
2  & J000552.33$-$000655.7& 5.850 & 23.09 & 20.50 & \ldots& 19.87 & \ldots& \ldots& \ldots& \ldots& -25.86 &   \citet{fan04} &  S82  \\
3  & J000825.77$-$062604.6& 5.929 & 22.85 & 20.35 & \ldots& 19.43 & \ldots& \ldots& 16.72 & 16.03 & -26.04 & \citet{jiang15} &    O  \\
4  &  J002806.57+045725.3 &  6.04 & 24.00 & 20.49 & 19.59 & 19.16 & 19.05 & 18.32 & \ldots& \ldots& -26.38 & \citet{jiang15} &    O  \\
5  &  J010013.02+280225.8 &  6.30 & 20.84 & 18.33 & \ldots& 17.00 & 15.98 & 15.20 & 14.45 & 13.63 & -29.10 &    \citet{wu15} &    M  \\
6  & J012958.51$-$003539.7& 5.779 & 24.48 & 22.13 & \ldots& 21.78 & \ldots& \ldots& \ldots& \ldots& -24.39 & \citet{jiang09} &  S82  \\
7  &  J014837.64+060020.0 & 5.923 & 22.25 & 19.31 & 18.91 & 18.37 & 17.72 & 17.13 & 15.90 & 15.09 & -27.08 & \citet{jiang15} &  M+O  \\
8  &  J020332.38+001229.4 &  5.72 & 23.76 & 20.75 & 19.85 & 19.05 & 17.75 & 17.32 & 16.35 & 16.06 & -25.74 & \citet{jiang08} &  S82  \\
9  & J023930.24$-$004505.3&  5.82 & 24.51 & 22.08 & 21.62 & 21.15 & \ldots& \ldots& \ldots& \ldots& -24.50 & \citet{jiang09} &  S82  \\
10 & J030331.41$-$001912.9& 6.078 & 24.17 & 20.97 & 20.60 & 20.44 & 19.78 & 18.95 & \ldots& \ldots& -25.31 & \citet{jiang08} &  S82  \\
11 &  J035349.73+010404.6 & 6.072 & 23.22 & 20.51 & 20.12 & 19.45 & 18.53 & 18.16 & \ldots& \ldots& -26.49 & \citet{jiang08} &  S82  \\
12 &  J081054.32+510540.1 &  5.80 & 21.52 & 19.34 & \ldots& 18.77 & \ldots& \ldots& 16.94 & 15.87 & -26.98 &      This paper &    M  \\
13 &  J081827.39+172251.8 & 6.02  & 22.62 & 19.67 & \ldots& 18.54 & \ldots& \ldots& \ldots& \ldots& -27.37 &  \citet{fan06a} &    M  \\
14 &  J083525.76+321752.6 &  5.89 & >23.0 & 20.73 & \ldots& 20.50 & \ldots& \ldots& \ldots& \ldots& -25.76 &      This paper &    O  \\
15 &  J083643.86+005453.2 & 5.810 & 20.97 & 18.71 & 18.27 & 17.70 & 17.02 & 16.18 & 15.23 & 14.46 & -27.86 &  \citet{fan01a} &    M  \\
16 &  J084035.09+562419.9 & 5.844 & 22.43 & 19.76 & \ldots& 19.00 & \ldots& \ldots& \ldots& \ldots& -26.64 &  \citet{fan06a} &    M  \\
17 &  J084119.52+290504.4 & 5.98  & 22.47 & 19.86 & 19.74 & 19.08 & 18.62 & 17.84 & \ldots& \ldots& -27.08 &  \citet{goto06} &    O  \\
18 &  J084229.43+121850.5 & 6.069 & 23.31 & 19.56 & \ldots& 18.84 & \ldots& \ldots& 15.76 & 15.42 & -26.85 & \citet{jiang15} &  M+O  \\
19 &  J085048.25+324647.9 & 5.867 & >22.5 & 19.95 & \ldots& 18.90 & \ldots& \ldots& 16.39 & 15.18 & -26.74 & \citet{jiang15} &    O  \\
20 &  J092721.82+200123.6 & 5.772 & 22.12 & 19.88 & \ldots& 19.01 & \ldots& \ldots& 16.66 & 15.68 & -26.78 &  \citet{fan06a} &    M  \\
21 &  J103027.09+052455.0 & 6.309 & 22.90 & 19.62 & 19.27 & 18.85 & 18.37 & 17.78 & 16.49 & 15.44 & -27.53 &  \citet{fan01a} &  M+O  \\
22 & J104433.04$-$012502.1& 5.778 & 21.68 & 19.07 & 18.87 & 18.31 & 17.92 & 17.03 & 16.36 & 15.56 & -27.61 &   \citet{fan00} &  M+O  \\
23 &  J104845.05+463718.4 & 6.198 & 22.43 & 19.85 & \ldots& 18.40 & \ldots& \ldots& 16.26 & 16.24 & -27.51 &   \citet{fan03} &    M  \\
24 &  J113717.72+354956.9 & 6.03  & 22.55 & 19.54 & \ldots& 18.41 & \ldots& \ldots& 16.29 & 15.78 & -27.08 &  \citet{fan06a} &    M  \\
25 &  J114338.34+380828.7 &  5.81 & 21.97 & 19.76 & \ldots& 18.98 & \ldots& \ldots& 16.93 & 16.03 & -26.76 &      This paper &    M  \\
26 &  J114803.28+070208.3 & 6.339 & 23.20 & 20.80 & 19.74 & 19.36 & 18.39 & 17.51 & 16.39 & 15.48 & -26.41 &      This paper &    M  \\
27 &  J114816.64+525150.3 & 6.419 & 23.18 & 19.98 & \ldots& 18.25 & \ldots& \ldots& 15.66 & 15.18 & -27.80 &   \citet{fan03} &    M  \\
28 &  J120737.43+063010.1 & 6.040 & >23.5 & 20.39 & 19.51 & 19.35 & \ldots& 17.50 & 16.53 & 14.82 & -26.60 & \citet{jiang15} &    O  \\
29 &  J124340.81+252923.9 &  5.85 & 23.08 & 20.22 & 19.81 & 19.21 & 18.29 & 17.54 & 16.70 & 15.52 & -26.22 &      This paper &    M  \\
30 &  J125051.93+313021.9 & 6.15  & 22.15 & 19.48 & 19.54 & 18.92 & 18.37 & 17.44 & \ldots& \ldots& -27.11 &  \citet{fan06a} &    M  \\
31 &  J125757.47+634937.2 &  6.02 & 23.50 & 20.60 & 20.39 & 19.78 & \ldots& \ldots& 16.71 & 16.48 & -26.14 & \citet{jiang15} &    O  \\
32 &  J130608.25+035626.3 & 6.016 & 22.35 & 19.29 & 19.24 & 18.86 & 18.69 & 17.34 & 15.99 & 15.52 & -27.32 &  \citet{fan01a} &    M  \\
33 &  J131911.29+095051.3 & 6.132 & 22.55 & 19.99 & 19.10 & 18.76 & \ldots& \ldots& 16.73 & 15.71 & -27.12 &  \citet{mort09} &    O  \\
34 &  J133550.80+353315.8 & 5.901 & 22.67 & 20.10 & 19.38 & 18.90 & \ldots& 17.61 & 16.81 & 16.13 & -26.81 &  \citet{fan06a} &    M  \\
35 &  J140319.13+090250.9 &  5.86 & 22.73 & 20.48 & 19.70 & 19.17 & 18.59 & 17.93 & 17.09 & 15.95 & -26.27 & \citet{jiang15} &    O  \\
36 &  J141111.27+121737.3 & 5.927 & 22.88 & 19.58 & 19.56 & 19.20 & 18.28 & 17.45 & 16.76 & 15.61 & -26.75 &   \citet{fan04} &  M+O  \\
37 &  J143611.73+500707.0 & 5.85  & 22.76 & 20.00 & \ldots& 19.04 & \ldots& \ldots& \ldots& \ldots& -26.51 &  \citet{fan06a} &    M  \\
38 &  J154552.08+602824.0 &  5.78 & 21.27 & 19.09 & \ldots& \ldots& \ldots& \ldots& 16.00 & 15.16 & -27.37 &  \citet{wang16a} &   M  \\
39 &  J160253.98+422824.9 & 6.09  & 22.88 & 19.81 & \ldots& 18.46 & \ldots& \ldots& 16.14 & 15.03 & -26.85 &   \citet{fan04} &    M  \\
40 &  J160937.27+304147.7 &  6.16 & >22.5 & 20.26 & 20.01 & 19.39 & 18.72 & 18.15 & 17.52 & 17.08 & -26.62 &      This paper &    M  \\
41 &  J162100.92+515548.8 &  5.71 & 21.86 & 19.70 & \ldots& 19.11 & \ldots& \ldots& 15.71 & 14.86 & -26.94 &      This paper &    M  \\
42 &  J162331.80+311200.6 & 6.247 & 24.50 & 19.67 & 19.72 & 19.16 & 18.45 & 17.86 & 16.85 & 15.44 & -27.04 &   \citet{fan04} &  M+O  \\
43 &  J163033.89+401209.7 & 6.058 & 23.28 & 20.34 & \ldots& 19.38 & \ldots& \ldots& \ldots& \ldots& -26.14 &   \citet{fan03} &    O  \\
44 &  J205321.77+004706.8 &  5.92 & 24.13 & 21.34 & \ldots& 20.46 & \ldots& \ldots& \ldots& \ldots& -25.54 & \citet{jiang09} &  S82  \\
45 & J205406.50$-$000514.4& 6.038 & 23.23 & 20.74 & \ldots& 19.18 & \ldots& \ldots& \ldots& \ldots& -26.09 & \citet{jiang08} &  S82  \\
46 & J211951.89$-$004020.1&  5.87 & 23.99 & 21.67 & \ldots& 20.87 & \ldots& \ldots& \ldots& \ldots& -24.73 &      This paper &  S82  \\
47 &  J214755.42+010755.5 &  5.81 & 24.21 & 21.61 & 20.92 & 20.79 & \ldots& \ldots& \ldots& \ldots& -25.00 & \citet{jiang09} &  S82  \\
48 &  J230735.36+003149.3 &  5.87 & 25.16 & 21.91 & 20.99 & 20.43 & \ldots& \ldots& \ldots& \ldots& -24.71 & \citet{jiang09} &  S82  \\
49 &  J231038.88+185519.7 & 6.003 & 21.66 & 19.21 & \ldots& 17.94 & \ldots& \ldots& 15.80 & 15.42 & -27.61 &      This paper &    M  \\
50 & J231546.58$-$002357.9& 6.117 & 23.80 & 20.85 & \ldots& 19.94 & \ldots& \ldots& \ldots& \ldots& -25.41 & \citet{jiang08} &  S82  \\
51 &  J232514.25+262847.6 &  5.77 & 21.62 & 19.42 & \ldots& \ldots& \ldots& \ldots& 16.19 & 15.41 & -26.98 &  \citet{wang16a} &    M  \\
52 &  J235651.58+002333.3 &  6.00 & 24.64 & 21.74 & \ldots& 21.18 & \ldots& \ldots& \ldots& \ldots& -24.84 & \citet{jiang09} &  S82  \\
\enddata
\tablenotetext{a}{The upper limits for four quasars indicate $3\sigma$ upper
limits.}
\tablenotetext{b}{`M': main survey, `O': overlap regions, `S82': Stripe 82, 
`M+O': main survey and overlap regions. The details are explained in 
Section 4.1.}
\end{deluxetable*}

\section{A sample of 52 SDSS quasars at $z\sim6$}

In this section we summarize our survey of $z\sim6$ quasars in the SDSS, and
present the final sample of 52 quasars discovered since 2000.
We then calculate the survey area coverage and the quasar sample completeness.
This information is used to derive the QLF at $z\sim6$ and the evolution
of luminous quasars at high redshift.

\subsection{The Quasar Sample}

Table 2 gives the basic data for the 52 SDSS quasars at $z\sim6$. 
They are ordered by R.A. Column 2 lists the quasar coordinates determined by 
the SDSS. Column 3 shows the quasar redshifts, taken from different resources, 
including the quasar discovery papers or follow-up observation papers.
The redshifts were mostly measured from emission lines, such as \lya\ in the 
(observed-frame) optical, \mgii\ in the near-IR 
\citep[e.g.][]{jiang07,kurk07,kurk09,mort09,derosa11}, or CO in the radio 
\citep[e.g.][]{car10,wang11,wang13}. Some quasars have very weak emission 
lines in the optical and near-IR (i.e., the rest-frame UV), and their 
redshifts were measured from the onset of the \lya\ absorption. 
In Column 2, the redshifts measured from \lya\ emission or absorption features 
are accurate to the second decimal place, and the redshifts measured from 
\mgii\ or CO lines are accurate to the third decimal place.
Columns 4--9 shows the photometry in the $izYJHK$ bands, and Columns 10--11 
shows the photometry in the first two Wide-field Infrared Survey Explorer 
\citep[WISE;][]{wri10} bands W1 and W2 at 
3.4 and 4.6 $\mu$m, respectively. The optical magnitudes are expressed in the 
AB system, and the near-IR and mid-IR magnitudes are in the Vega system.
Column 12 is the absolute AB magnitude of the continuum at rest-frame 1450 \AA\ 
($M_{1450}$). We have converted all published $M_{1450}$ values to the 
cosmology used in this paper. 
Column 13 shows the references of the quasar discovery papers.
Column 14 indicates if a quasar is in the SDSS main survey (`M'), overlap 
region (`O'), or Stripe 82 (`S82'). Note that seven main-survey quasars are 
also located in overlap regions, and they are marked as `M+O'.

The majority of the 52 quasars were found by Fan et al. (2000--2006) and Jiang 
et al. (2008, 2009, 2015, and this paper). Three quasars in the sample, 
J0100+2802 ($z=6.30$), J1545+6028 ($z=5.78$), and J2325+2628 ($z=5.77$), were 
reported by \citet{wu15} and \citet{wang16a}. They were discovered using 
combination of the SDSS and WISE imaging data \citep[see also][]{bla13,yan13}, 
and were also selected by our standard selection criteria mentioned in 
Section 2. We thus included these three quasars in our final sample.
We also included J0841+2905 ($z=5.98$) found by \citet{goto06} and J1319+0950 
($z=6.132$) found by \citet{mort09}. These two quasars do not meet our 
selection cut $\sigma_z<0.1$ mag in the SDSS single-epoch images, but they are 
located in the overlap regions, and were recovered as overlap-region quasars 
by \citet{jiang15}. 

We did not include three quasars reported by \citet{cool06}, \citet{mcg06}, 
and \citet{wang16a}. \citet{wang16a} presented three $z\sim6$ quasars, and two 
of them meet our selection criteria as mentioned above. The third one was not 
selected by us, because its $\sigma_z$ is slightly larger than 0.1 mag.
We did not include the quasar (at $z=5.85$) of \citet{cool06}, which is 
significantly fainter than our selection limit in the SDSS images. 
We did not include the radio-loud quasar (at $z=6.12$) of \citet{mcg06}, which
was found from its radio emission. It is bright in the optical, but it is 
strongly blended with a much brighter neighbor in the SDSS images, and was not 
separately detected by the SDSS pipeline. We also did not include the 
radio-loud quasar (at $z=5.95$) of \citet{zei11} found in Stripe 82, which is 
fainter than our selection limit in the co-added Stripe 82 images.

\begin{figure}
\epsscale{1.2}
\plotone{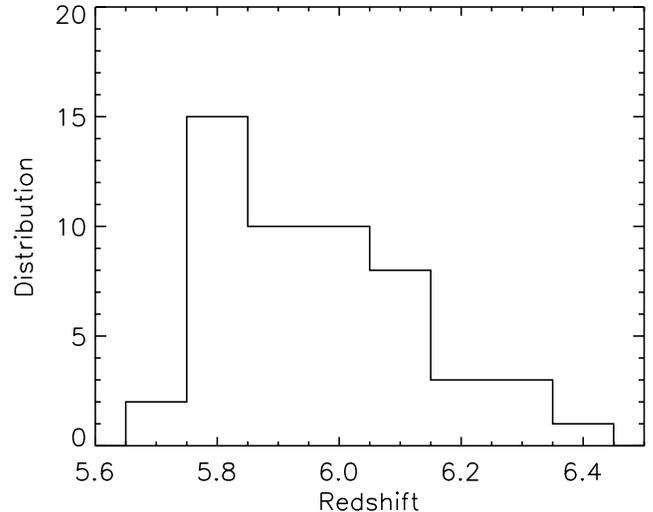}
\caption{Redshift distribution of the 52 SDSS quasars.
The number of quasars decreases rapidly from $z\sim5.8$ to $z\sim6.4$.
\label{fig:redshift}}
\end{figure}

Figure \ref{fig:redshift} shows the redshift distribution of the 52 SDSS 
quasars. The number of quasars above our flux limits decreases rapidly from 
$z\sim5.8$ to $z\sim6.4$. Figure \ref{fig:allspec} in the Appendix shows their 
optical spectra. Most of the spectra were taken from the quasar discovery 
papers listed in Column 13 of Table 2. The spectrum of J1319+0950 was 
presented in \citet{mcg15}. The spectrum of J0841+2905 was obtained from the 
MMT. Note that there exist higher S/N optical spectra for some quasars
\citep[e.g.][]{bec11,bec15} that are not shown in Figure \ref{fig:allspec}.

Among the 52 quasars, 47 quasars belong to one of three statistically complete
samples: the main survey sample, the overlap region sample, and the Stripe 82
sample. There are 24 quasars in the SDSS main survey, 17 in the overlap
regions (7 of them also belong to the main survey), and 13 in Stripe 82.
The other 5 quasars are beyond our standard selection criteria and not part of
the complete samples. They are J1335+3533 and J1436+5007 from \citet{fan06a}, 
and J1148+0702, J1609+3041, and J1621+5155 in this paper.
We will now derive the QLF based on the 47 quasars.

\subsection{Area Coverage}

In this subsection, we calculate the effective area coverage for our quasar 
samples. The calculation of effective area is not straightforward for several 
reasons, including the SDSS imaging survey geometry, possible missing data, 
existence of very bright stars (resulting in large `holes' in object 
catalogs), and so on. We estimate the effective area using the Hierarchical 
Equal Area isoLatitude Pixelization \citep[HEALPix;][]{gor05}.
HEALPix hierarchically tessellates the spherical sky into a mesh of 
curvilinear quadrilaterals. The base resolution, or the lowest resolution
level (we call it Level 0), consists of 12 pixels over the celestial sphere.
The resolution level increases by dividing each pixel into 4 subpixels with 
identical area. At level higher than 1, each pixel has 8 neighboring pixels, 
except for 24 pixels (each of them has 7 neighboring pixels).
The effective area of our samples is calculated by 
adding up the area of the pixels that cover our data points. Obviously, the 
accuracy depends on the resolution level of the HEALPix pixels and the spatial 
density of data points. 

Our data points are drawn from the SDSS Query CasJobs online server. 
We use all SDSS `Primary' objects with $r<22.5$ mag and $i<22.5$ mag for the 
main survey. We only remove a tiny fraction of objects with the SDSS 
processing flags `BRIGHT', `EDGE', and `SATUR'. For the overlap regions, we 
use all `Primary' and 'Secondary' objects down to $r=23.0$ mag and $i=23.0$ 
mag. The average density of the data points is about 5 objects per square 
arcminute. The data points for Stripe 82 are drawn from our stacked 
images/catalogs \citep{jiang14}, and the number density is much higher than 
that for the other two regions.

\begin{figure}
\epsscale{1.2}
\plotone{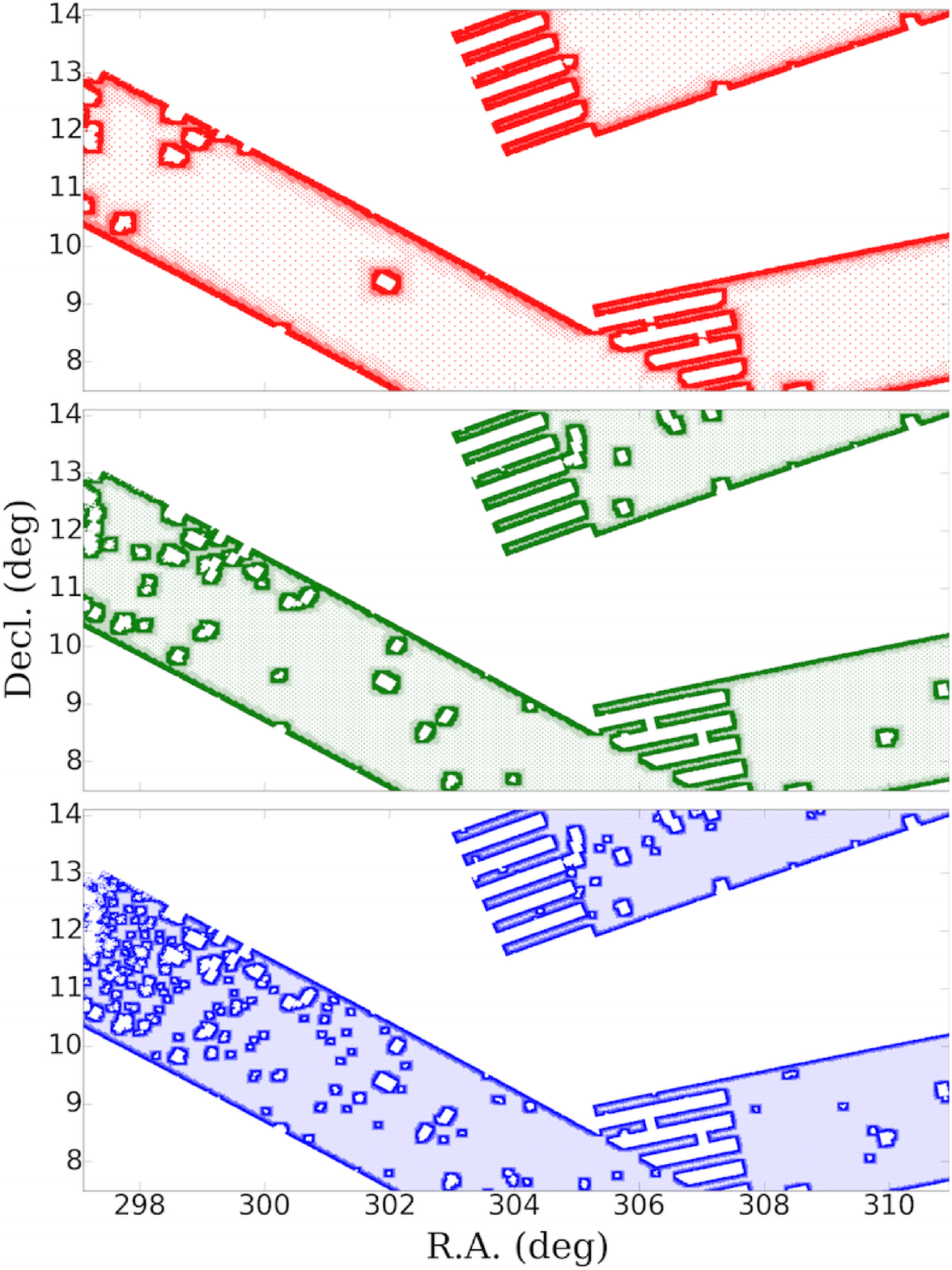}
\caption{An example of effective coverage maps (for the same regions) measured 
by HEALPix with three different starting resolution levels, Levels 9 
(top panel), 10 (middle panel), and 11 (bottom panel). We use Level 10 as our
starting resolution level (see the main text for the details).
\label{fig:coverage}}
\end{figure}

For a given dataset, the starting resolution level is critical for area 
calculation. We have tried different starting resolution levels, and found 
that HEALPix Level 10 (i.e., the HEALPix base resolution is divided 10 times)
is the best for the SDSS. Figure \ref{fig:coverage} shows an example that 
compares the results from three different starting resolution levels, Levels 9 
(top panel), 10 (middle panel), and 11 (bottom panel). 
The added `holes' from Level 9 to Level 10 are almost all real,
primarily caused by missing data and very bright stars. On the contrary,
the majority of the added `holes' from Level 10 to Level 11 are not real, 
but represent relatively empty regions of the sky in the SDSS imaging. 
Therefore, we use HEALPix Level 10 as our base resolution or starting 
resolution, which is 11.8 square arcminutes per pixel.

Now we classify all pixels into three categories. In the first category are 
the pixels that do not cover any data points, and these regions are beyond the
effective coverage of our survey. The pixels in the second category (hereafter 
boundary pixels) are the close neighbors to the pixels in the first category, 
i.e., each boundary pixel has at least one neighboring pixel in the first 
category. The boundary pixels include outer boundaries and inner boundaries 
(the edges of the inner `holes', see Figure \ref{fig:coverage}). The third 
category contains all remaining pixels (hereafter non-boundary pixels). 
All non-boundary pixels at Level 10 constitute the major part of the total 
effective coverage. 

The accuracy of area calculation is now determined by boundary pixels. 
We refine boundary pixels by gradually increasing the resolution level, until 
the resolution roughly matches the average surface density of the data points. 
For the main survey and overlap regions, the boundary pixels are calculated to 
Level 13, at which the resolution is 0.18 square arcminutes per pixel 
or 5.5 pixels per square arcminute, matching the density of 5 objects
per square arcminute. For Stripe 82, the boundary pixels are calculated to 
Level 14, at which the resolution is 0.046 square arcminutes per pixel.

The effective area of the main survey is $11,240\pm59$ deg$^2$. 
The uncertainty quoted here is the contribution of the boundary pixels.
As we mentioned 
earlier, we mainly focused on high galactic latitude $|b|>30$ deg (excluding 
$|\rm Decl.|<1.3$), which has an area of 10,371 deg$^2$. We also include the 
lower galactic latitude region between 20 and 30 deg used by \citet{fan06a}. 
These SDSS images were taken before 2005 June, and their area is 
about 869 deg$^2$. The main survey covers 24 luminous quasars at $z\sim6$, and 
the spatial density of these quasars is very low (about 1 per 468 deg$^2$).

The effective area of the overlap regions is $4223\pm139$ deg$^2$. The 
uncertainty, or the contribution of the boundary pixels, is relatively large, 
due to the more complex geometry of the overlap regions. For the overlap 
regions, we only considered high galactic latitude $|b|>30$ deg. A total of 17 
quasars fall in the overlap regions, and 7 of them also belong to the main 
survey. In other words, 10 quasars are fainter than a $10\sigma$ detection in 
the SDSS single-epoch $z$-band images. 

The effective area of Stripe 82 is $277\pm1$ deg$^2$.
For Stripe 82, we did not use the region of $\rm RA<310$ deg. This region is 
at relatively low galactic latitude ($|b|<24$). In addition, the coadded 
images in this region are significantly shallower than other regions, due to 
the smaller number of imaging runs covering this region. We found 13 quasars 
in Stripe 82. The spatial density is about 1 per 21 deg$^2$, which is much
higher than the density in the main survey.

\subsection{Sample Completeness}

We use simulations to estimate the completeness of the quasar sample.
The incompleteness comes from our quasar selection criteria, i.e., the
color cuts and survey limits that we applied in Section 2. 
We describe the completeness as a selection function, the probability that a 
quasar with a given magnitude ($M_{1450}$), redshift ($z$), and intrinsic 
spectral energy distribution (SED) meets our selection criteria.
We generate a grid of model quasars using the simulations by \citet{mcg13}, 
which is an updated version of the simulations by \citet{fan99}.
The model SEDs are designed to reproduce the colors of $\sim$60,000 quasars
at $2.2<z<3.5$ from the SDSS BOSS survey \citep{ross12}.
Each model SED consists of a broken power-law continuum, a series of emission 
lines with Gaussian profiles, and a scaled Fe emission template. 
The distributions of spectral features such as the continuum spectral index, 
line EW, and line FWHM match those from the BOSS quasars. The model also 
incorporates the relations between spectral features and quasar luminosity, 
such as the Baldwin effect and blueshifted lines (see \citet{mcg13} for 
details). The model does not take into account broad-absorption-line (BAL) 
quasars and quasars with weak emission lines \citep{plo15}. These quasars
have slightly different colors, but the overall impact on our calculation
is negligible compared to the statistical uncertainties derived in the next
subsections.

We extend the model to higher redshifts under the assumption that the shape
of the quasar SED does not evolve with redshift. The only significant 
difference is the increasing neutral H absorption in \lya\ forests towards
higher redshifts. Finally, photometry is derived from the SED models and
photometric errors appropriate for each survey region are added. We draw a 
large representative sample of objects from the SDSS archive (or from our 
stacked images/catalogs for Stripe 82). From this sample, we obtain the
relation between magnitude and error in the $i$ and $z$ bands in a 2D space,
giving us an error distribution at each magnitude. The errors are added 
to the model quasars so that the error distributions match those from the 
real data above. The $J$-band errors are added in the same way so that the
error distributions for the model quasars match those from our $J$-band data.

\begin{figure}
\epsscale{1.2}
\plotone{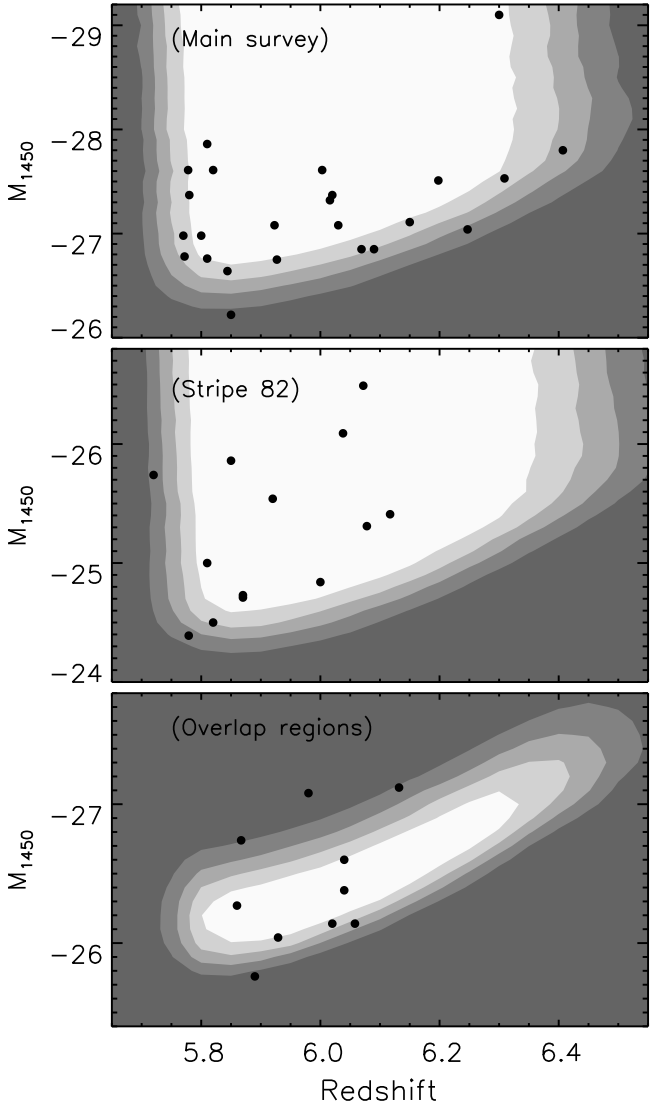}
\caption{Quasar selection function as a function of $M_{1450}$ and $z$ for the
main survey sample (top panel), the Stripe 82 sample (middle panel), and
the overlap-region sample (bottom panel). The contours in the top and middle
panels are selection probabilities from 0.8 to 0.2 with an interval of 0.2. 
The contours in the bottom panel are probabilities from 0.6 to 0.15 with an 
interval of 0.15. 
The filled circles indicate the locations of the quasars in the three samples.
\label{fig:completeness}}
\end{figure}

As we did in \citet{fan01a} and \citet{jiang08}, we compute the average 
selection probability, $p(M_{1450},z)$, 
as a function of $M_{1450}$ and $z$ after photometric 
errors are properly incorporated. Figure \ref{fig:completeness} shows the 
selection function for the main survey sample (top panel), the Stripe 82 
sample (middle panel), and the overlap-region sample (bottom panel).
The filled circles indicate the locations of the quasars in the two samples. 
The probability decline at $z<5.8$ and $z>6.3$ is caused by the color cuts on 
the $i_{AB}-z_{AB}$ and $z_{AB}-J$ colors, and the probability decline at the 
low luminosity end is due to the survey limit in the $z$ band.
In the top panel, one quasar (J1243+2529 discovered in this paper) has a 
probability below 20\%. The reason is that this quasar is relatively
faint with $z_{AB}=20.22$ mag, below the nominal limit of $z_{AB}\approx20.0$ 
mag for $\sigma_z=0.10$ mag in the SDSS. But its $z$-band error 
($\sigma_z=0.10$ mag) still satisfies our selection criteria.
In the middle panel, the $z=5.72$ quasar (J0203+0012) has the lowest 
probability. This quasar was originally found to be at $z=5.85$, and it 
appears to be a $z=5.85$ quasar in \ref{fig:allspec}. It was later confirmed 
to be a BAL quasar at $z=5.72$, and its \lya\ emission has been largely 
absorbed \citep{mort09}. In the bottom panel, the contours are different from 
those for the other two samples. This is because the 10 quasars in the overlap 
regions were selected in a small magnitude range, i.e., $7\sigma$ to 
$10\sigma$ detections in the SDSS $z$-band images.
The three brightest quasars in this panel are brighter than the faintest
quasars in the top panel. However, they are located in regions that are much 
shallower than the nominal depth of the SDSS single-epoch images, and thus 
were not selected in our main survey.

In Figure \ref{fig:colorselect}, we plot the SDSS quasars on the $z-J$ versus 
$i-z$ color-color diagram. The figure shows that the $z-J$ colors of the SDSS 
quasars are on average bluer than those of the simulated quasars (black dots). 
We check $\sim$10 quasars with the bluest $z-J$ colors in the figure, and find 
that most of them have very strong \lya\ emission that the model does not 
fully account for. Another likely reason is that $z\sim6$ quasars tend to have 
bluer rest-frame UV continuum colors compared to quasars at $2.2<z<3.5$. 
Near-IR observations of a large sample of $z\sim6$ quasars are needed to 
confirm this hypothesis. Nevertheless, quasars with blue $z-J$ colors are well 
within our selection criteria, and have little impact on our sample 
completeness.

\subsection{Binned Luminosity Function at $z\sim6$}

We first derive the binned luminosity function from the three subsamples
separately: the main survey sample with 24 quasars, the overlap region sample 
with 10 quasars, and the Stripe 82 sample with 13 quasars. The main survey and
Stripe 82 samples are divided into 4 and 2 discrete luminosity bins,
respectively. Redshift evolution is not considered here, but will be
taken into account when we parametrize the QLF later.
The volume densities of the quasars are calculated using the traditional
$1/V_{a}$ method, with the selection function included.

\begin{figure}
\epsscale{1.2}
\plotone{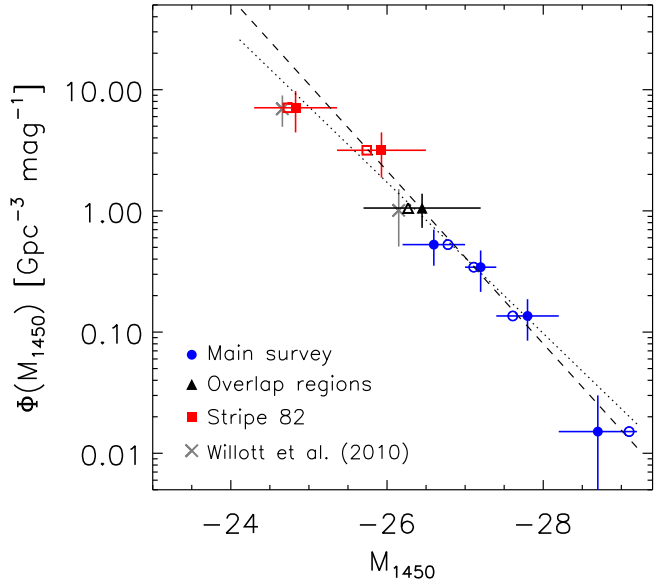}
\caption{Binned luminosity function for the SDSS quasars at $z\sim6$.
The filled symbols with error bars represent the binned QLF for the main
survey sample (blue circles), overlap region sample (black triangles), and 
Stripe 82 sample (red squares). The horizontal locations of the filled symbols 
indicate the centers of the luminosity bins, and the horizontal bars indicate 
the luminosity ranges 
that the bins cover. The horizontal locations of the open symbols indicate
the median luminosity values in individual bins.
The crosses show the results from the CFHQS \citep{wil10}, and they are
consistent with the SDSS results. The dotted line is a power-law fit to all
data points, and the dashed line is a power-law fit to all but the faintest
data point.
\label{fig:lf}}
\end{figure}

The binned QLF is shown in Figure \ref{fig:lf}. The horizontal locations of 
the filled symbols represent the centers of the luminosity bins, and the 
horizontal bars indicate the luminosity ranges that the bins cover. 
In the main survey sample, the $z=6.30$ quasar of \citet{wu15} is much 
brighter than the others, and is thus put in one luminosity bin. The other 
three bins have similar numbers of quasars.
We show the median value of the quasar luminosities in each bin as the open 
symbols in the figure. The results are consistent with our previous results 
based on smaller samples, e.g., Figure 6 in \citet{jiang08} and Figure 3 in 
\citet{jiang09}. They also agree with the results from the CFHQS 
\citep[gray crosses in the figure; see][]{wil10}.

The binned QLF from our SDSS sample in \citet{jiang08,jiang09} is well fit by 
a single power law $\Phi(L_{1450})\propto L_{1450}^{\beta}$, or,
\begin{equation} \label{eq:spl}
\Phi(M_{1450})=\Phi^{\ast}10^{-0.4(\beta+1)(M_{1450}+26)},
\end{equation}
with a steep slope $\beta$ around --2.9. Figure \ref{fig:lf} shows that the 
updated binned QLF can also be described as a power law. The best fit (the 
dotted line in Figure \ref{fig:lf}) to all SDSS data points results in a slope 
of $\beta=-2.55\pm0.17$. The fit is dependent on the 
luminosity bin sizes that we chose. In order to remove this dependence, we use 
a maximum likelihood analysis to find the best fit. The likelihood function 
\citep[e.g.][]{mar83} is written as
\begin{multline} \label{eq:likelihood}
S = -2\sum \ln[\Phi(M_{i},z_{i})p(M_{i},z_{i})] \\
	+ 2\int_{\Delta M}\int_{\Delta z} \Phi(M,z)p(M,z) \frac{dV}{dz} dzdM,
\end{multline}
where the sum is over all quasars in the sample, and the integral is over the
full luminosity and redshift space of the sample. The best fit is
$\beta=-2.56\pm0.16$, consistent with the above result.

At low redshift, QLFs are commonly characterized using a double power law,
\begin{multline} \label{eq:dpl}
\Phi(M,z) \\
	=\frac{\Phi^{\ast}(z)} 
   {10^{0.4(\alpha+1)(M-M^{\ast}(z))}+10^{0.4(\beta+1)(M-M^{\ast}(z))}},
\end{multline}
where $M^{\ast}(z)$ is the characteristic magnitude, and $\alpha$ is the slope 
at the faint end. Our SDSS data are apparently not deep enough to reach 
$M^{\ast}$. However, the recent discovery of much fainter high-redshift 
quasars \citep[e.g.][]{wil10,kas15,kim15,mat16} have suggested a flatter 
faint-end slope, and the faintest data point in our sample may be affected by
the turnover. In order to better constrain the bright-end slope $\beta$, 
we also fit a single power law to all SDSS density points but the faintest one 
in Figure \ref{fig:lf}, i.e., the luminosity range $M_{1450}<-25.3$ mag. 
The best fit is $\beta=-2.78\pm0.24$. This is slightly steeper than the slopes 
found above from the fit to all data points, but is consistent with the 
value $\beta=-2.81$ reported by \citet{wil10}. We thus conclude that the 
bright-end slope of the $z\sim6$ QLF is around $\beta=-2.8$, and we will
adopt $\beta=-2.8\pm0.2$ in the rest of the paper.

\subsection{Double power law fit to the z$\sim$6 QLF}

\begin{figure}
\epsscale{1.2}
\plotone{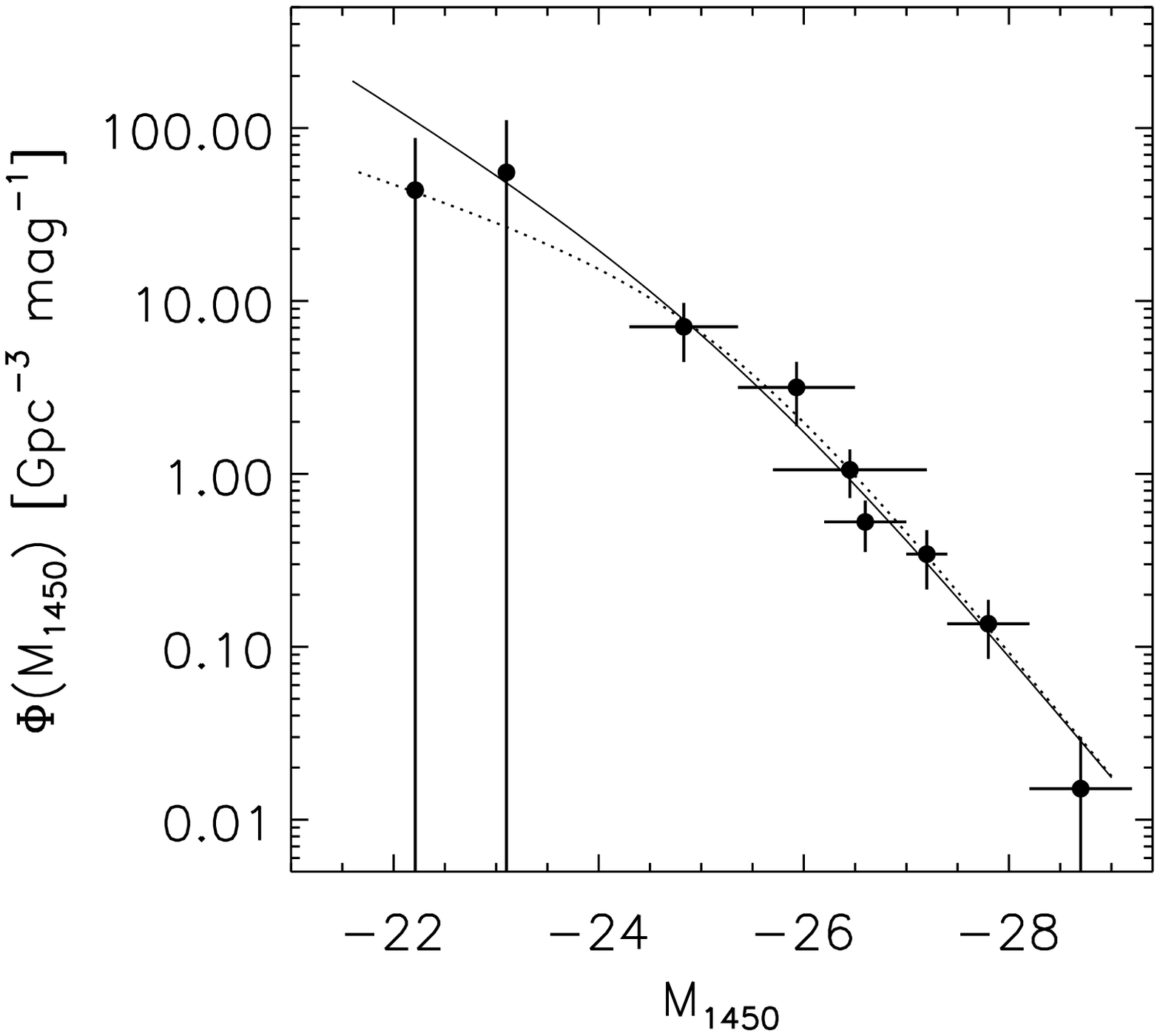}
\caption{QLF at $z\sim6$ fitted by a double power law.
The two faintest data points represent the two quasars from \citet{wil10} and 
\citet{kas15}, respectively. The other data points are the binned SDSS
luminosity function from Figure \ref{fig:lf}. The solid line is the best 
double power law (Equation \ref{eq:dpl}) fit using the maximum likelihood 
method. The dotted line represents the QLF with a fixed slope $\alpha=-1.5$
derived by \citet{wil10}. It is consistent with our QLF at the bright end. 
The faint end of the QLF is poorly constrained.
\label{fig:lf2}}
\end{figure}

We now parametrize the double power law QLF (Equation \ref{eq:dpl}) at 
$z\sim6$. In order to constrain the slope $\alpha$, faint-end data points are
required. However, only a small number of $z\sim6$ quasars discovered so far 
are fainter than $M_{1450}\sim-24$ mag \citep[e.g.][]{wil10,kas15,kim15,mat16}. 
In our analysis, we include two faint quasars from \citet{wil10} and 
\citet{kas15}. 
They are represented as the two faintest data points in Figure 
\ref{fig:lf2}. Their effective area coverage and sample completeness were
carefully derived in the above papers, and have been incorporated into our
calculation. We do not include an object in \citet{kas15} with a narrow 
\lya\ line. It is likely a \lya-emitter, not a type 1 quasar or AGN.

The combined sample is still not sufficient to simultaneously constrain 
all parameters in Equation \ref{eq:dpl}, thus we choose to fix some of the 
parameters. We assume that $M^{\ast}(z)$ is constant over our 
redshift range, i.e., $M^{\ast}(z)=M^{\ast}$. The steep decline of the quasar 
density at high redshift can be described as \citep[e.g.][]{fan01b,mcg13},
\begin{equation} \label{eq:denEV}
\Phi^{\ast}(z)=\Phi^{\ast}(z=6)\,10^{k(z-6)}.
\end{equation}
Here we assume $k=-0.7$, derived from the density evolution of luminous 
quasars from $z\sim5$ to 6 (see details in Section 5.1). Furthermore, 
we fix the value of $\beta$ to be $-2.8$, as measured in Section 4.4. 
We then estimate the faint-end slope $\alpha$ and the characteristic magnitude
$M^{\ast}$ by applying a maximum likelihood analysis to Equation \ref{eq:dpl}. 
The results are $\alpha=-1.90_{-0.44}^{+0.58}$ and 
$M^{\ast}=-25.2_{-3.8}^{+1.2}$ mag.
The resultant $\Phi^{\ast}(z=6)$ from the best fit is 9.93 Gpc$^{-3}$ 
mag$^{-1}$. Thus the best-fit QLF at $z\sim6$ can be written as,
\begin{multline} \label{eq:dplfit}
\Phi(M,z) \\
   =\frac{9.93\times10^{-0.7(z-6)}} 
   {10^{0.4(-1.9+1)(M+25.2)}+10^{0.4(-2.8+1)(M+25.2)}},
\end{multline}
in units of Gpc$^{-3}$ mag$^{-1}$. Note that this is the observed QLF and 
does not take into account quasar intrinsic properties such as 
anisotropic emission and dust extinction \citep[e.g.][]{dip14}.

We perform two-dimensional Kolmogorov-Smirnov (K-S) tests to assess the
derived QLF. We generate a large sample ($>10,000$) of quasars drawn from
the derived QLF (Equation \ref{eq:dplfit}), covering the full
$M_{1450}$-redshift space shown in Figure \ref{fig:completeness}. The sample 
is then convolved with each of the three selection functions to produce
three samples of simulated objects. The resultant samples are compared with 
the three 
observed quasar samples using the K-S test \citep[e.g.][]{pea83,fas87}.
The probability found in each of the three cases is greater than 0.2, which
means that the hypothesis that two data sets are not significantly different 
is certainly correct. This indicates that the maximum likelihood analysis that
we did above is reasonable.

\begin{figure}
\epsscale{1.2}
\plotone{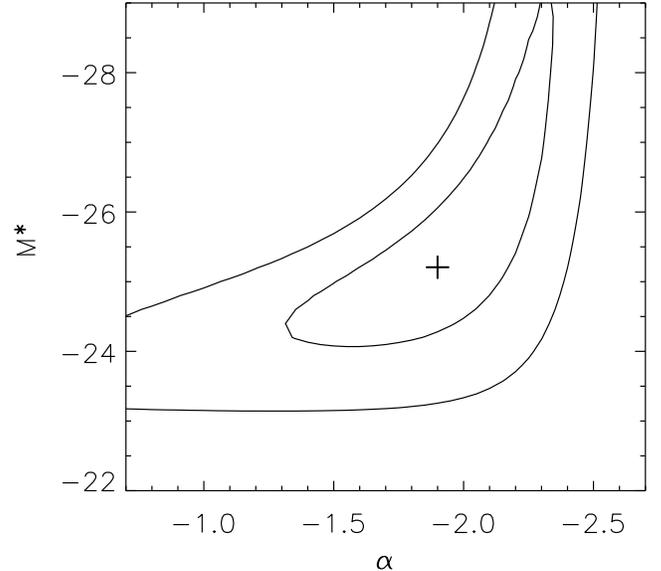}
\caption{Contours for the variation of the likelihood function with
$\alpha$ and $M^{\ast}$ in the region of the minimum $\chi^2$. The plus sign
indicates the position of the best fit. The contours represent the 68.3\%
(inner) and 95.4\% (outer) confidence regions. The two parameters are
correlated.
\label{fig:alpha_mstar}}
\end{figure}

The two free parameters $\alpha$ and $M^{\ast}$ are poorly constrained as 
Figure \ref{fig:alpha_mstar} shows; the uncertainties are due to the small 
number (2) of quasars at the faint end and the degeneracy between $\alpha$ and 
$M^{\ast}$. The real uncertainties are likely to be larger; by fixing $k$ 
and $\beta$ we have not accounted for the uncertainties in those parameters. 
At low redshift ($z\le3$), the QLF has a very steep 
bright-end slope $\beta\le-3$ and a much flatter faint-end slope 
$\alpha\sim-1.5$ \citep[e.g.][]{ric06,cro09,ross13}.
The bright-end slope at $z\ge4$ is found to be quite steep ($\beta\le-3$) as 
well \citep[e.g.][]{mcg13,yang16}. The steep bright-end slope $\beta=-2.8$ at 
$z\sim6$ does not evolve much from those at relatively lower redshifts.
On the other hand, the faint-end slope of the optical QLF at $z>3$ has not 
been well constrained \citep[e.g.][]{gli11,ike11,ike12,fio12}. 
Previous studies of the QLF at $z\sim6$ usually assumed a fixed slope 
$\alpha=-1.5$ as found for low-redshift quasars. For example, 
\citet{wil10} fixed $\alpha=-1.5$ and found $\beta=-2.81$. Their QLF is
consistent with ours at the bright end (Figure \ref{fig:lf2}).
Our results suggest that the 
faint-end slope at $z\sim6$ may be marginally steeper than --1.5 based on the 
faintest data points shown in Figure \ref{fig:lf2}, as already pointed out by 
\citet{kas15} and \citet{mat16}. However, 
the large uncertainties on $\alpha$ and $M^{\ast}$ will not be reduced
before a sizable faint quasar sample is obtained.

\section{Discussion}

\subsection{Density Evolution of Luminous Quasars at High Redshift}

\begin{figure}
\epsscale{1.2}
\plotone{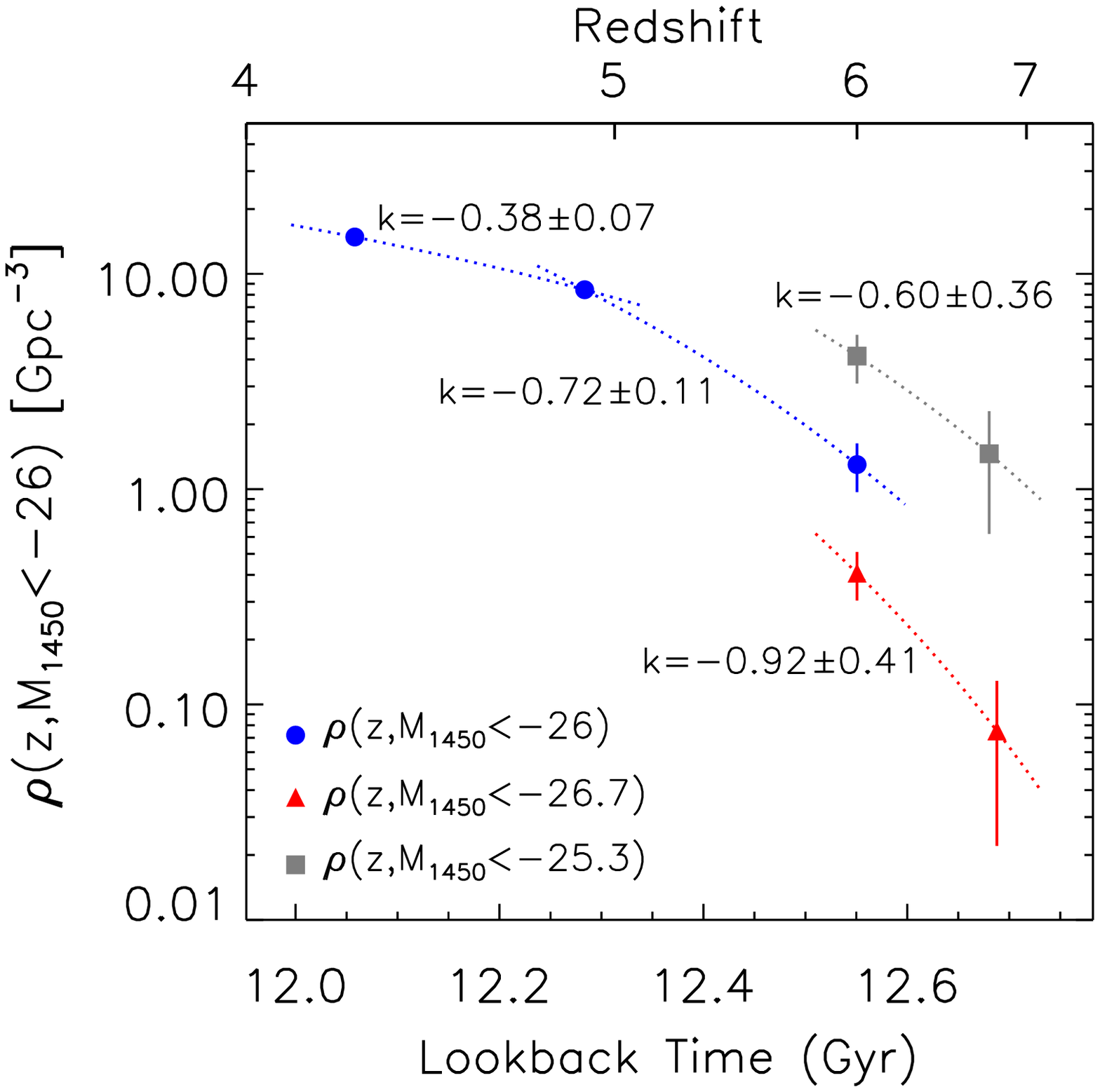}
\caption{Density evolution of luminous quasars at $z>4$. The blue circles, 
red triangles, and gray squares are the cumulative densities down to 
$M_{1450}=-26.0$, --26.7,
and --25.3 mag, respectively. The data points at $z\sim4$ and 5 are from 
\citet{mcg13}, the data points at $z=6$ are calculated from our new sample.
The gray data point at $z\sim7$ is taken from \citet{ven13}, and the red data 
point at $z\sim7$ is estimated from two $z>6.5$ quasars in the UKIDSS area
(see details in Section 5.1). The dotted lines are the power-law 
(Equation \ref{eq:denEV}) fits to
the data points. The figure shows a rapid decline of the
quasar spatial density from $z\sim5$ towards higher redshifts.
\label{fig:qsoev}}
\end{figure}

In Section 4.5, we described the quasar density evolution using Equation
\ref{eq:denEV}, where $k=-0.7$. Here we explore in detail the density 
evolution of luminous quasars at $z\ge4$. We work with the integral of the 
luminosity function from some fiducial lower luminosity $M$ to infinity,
\begin{equation} \label{eq:cQLF}
   \rho(<M,z) = \int_{-\infty}^{M}\,\Phi(M',z)\,dM'.
\end{equation}
In Figure \ref{fig:qsoev}, the blue circles show the results at $z\sim6$ for 
$M_{1450}=-26.0$ mag, together with the results at $z\sim4$ and 5 from 
\citet{mcg13}. The slopes of the lines give $k=-0.38\pm0.07$ from $z=4$ to 5, 
and $k=-0.72\pm0.11$ from $z=5$ to 6. \citet{fan01b} found $k=-0.47$ for 
quasars brighter than $M_B=-26$ mag at $3.5<z<5$. This value ($k=-0.47$) has 
been frequently used in more recent papers 
\citep[e.g.][]{wil10,mcg13}. \citet{mcg13} noticed that the evolution from 
$z\sim5$ to $\sim6$ is stronger for quasars brighter than $M_{1450}=-26$ mag, 
with $k=-0.7$. We update the density at $z=6$ and find $k=-0.72\pm0.11$ for 
the same redshift range, which is the same as the value given by 
\citet{mcg13}.

Alternatively, we add the density evolution $10^{k(z-6)}$ to 
Equation \ref{eq:spl}, and use the maximum likelihood analysis to find
the best fit to a single power law. The result is $k=-0.5\pm0.4$, consistent
with the above result. The large uncertainty is due to the limited sample 
size, short redshift baseline, and degeneracy between $k$ and $\beta$. 
Therefore, we chose to use $k=-0.72\pm0.11$ shown in Figure \ref{fig:qsoev}.
The underlying assumption is
that the density evolution at $z\sim6$ is similar to that at $z=5\sim6$.

Figure \ref{fig:qsoev} shows that the spatial density of luminous quasars from 
$z\sim5$ to 6 drops faster than that from $z\sim4$ to 5 
\citep[see also][]{mcg13}. This trend seems to continue towards higher 
redshift. So far the QLF at $z>6.4$ has not been well explored. We estimate 
the spatial density of luminous quasars at $z\sim7$ as follows. The UKIDSS 
team reported a 
$z=7.08$ quasar with $M_{1450}=-26.6\pm0.1$ mag \citep{mort11}, and a $z=6.53$ 
quasar with $M_{1450}=-27.4$ mag \citep{ven15}. The average effective area is 
roughly 3370 deg$^2$ for the two quasars (private communication with 
D.J. Mortlock and S.J. Warren). In order to include the $z=7.08$ quasar, we 
integrate Equation \ref{eq:cQLF} down to $M_{1450}=-26.7$ mag (instead of 
$M_{1450}=-26.0$ mag), which is roughly the limit of their quasar selection. 
Since the selection function for the two quasars has not 
been calculated, we assume that their selection probability is 1, which would 
slightly underestimate the density. Given the large statistical uncertainty,
the assumption is reasonable for the luminosity regime considered here 
(brighter than $M_{1450}=-26.7$ mag). The results are shown in Figure 
\ref{fig:qsoev}. The slope $k$ derived for the density evolution between
$z\sim6$ and 7 is $k=-0.92\pm0.41$. 

In addition, we estimate the density evolution of less luminous quasars
from $z=6$ to $\sim7$, using three quasars with $-26.0<M_{1450}<-25.3$ from
the VISTA VIKING survey \citep{ven13}. The cumulative density is 
integrated down to $M_{1450}=-25.3$ mag, which is roughly their quasar survey 
limit. As shown in Figure \ref{fig:qsoev}, the slope $k$ is $-0.60\pm0.36$. 
The above estimate on the density evolution at $z>6$ is rough, based on
two very small samples. Nevertheless, the trends seen in 
Figure \ref{fig:qsoev} suggest a rapid density decline of luminous quasars
from $z\sim5$ towards higher redshifts.

\subsection{Quasar Contribution to Reionization}

\begin{figure}
\epsscale{1.2}
\plotone{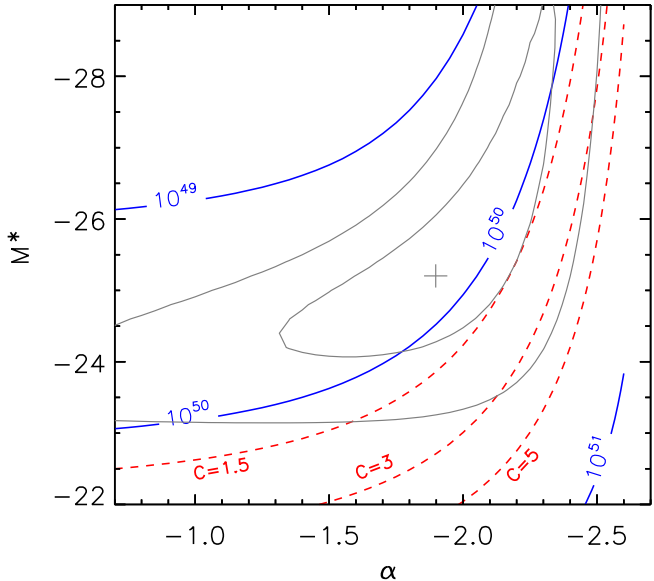}
\caption{Ionizing photon emissivity from quasars as a function of $\alpha$ and 
$M_{1450}^\ast$. The blue solid lines represent the total photon emissivity 
from quasars in units of Mpc$^{-3}$ s$^{-1}$. The red dashed lines represent 
the photon emissivity required to ionize the IGM at $z\sim6$ for the clumping 
factors $C=1.5$, $C=3$, and $C=5$. The gray plus sign and contours are the 
best-fit $\alpha$ and $M^{\ast}$ and their confidence regions from Figure 
\ref{fig:alpha_mstar}. 
\label{fig:reion}}
\end{figure}

We estimate the quasar contribution to the ionizing background at $z\sim6$.
We first calculate the number of ionizing photons provided by quasars based
on the QLF derived in Section 4. We assume a broken power-law quasar SED
as follows \citep{lus15},
\begin{equation}
	L_{\nu} \propto \left\{ \begin{array}{ll}
                 \nu^{-0.6}, & \mbox{if $ \lambda >$ 912 \AA;} \\
                 \nu^{-1.7}, & \mbox{if $ \lambda <$ 912 \AA.}
                        \end{array}
	\right.
\end{equation}
The spectral index may vary with quasar luminosity and background ionization 
rate \citep[e.g.][]{wyi11}, but we do not consider these complications here.
We integrate the SED over an energy range of 1--4 ryd (photons above 4 ryd
are consumed by \heii), and integrate the QLF over a luminosity range from 
$M_{1450}=-30$ to --18 mag. In Figure \ref{fig:reion}, the blue contours show 
the computed number of ionizing photons in units of Mpc$^{-3}$ s$^{-1}$ as a 
function of $\alpha$ and $M^{\ast}$. 

The number of ionizing photons is related to the minimum luminosity that we 
use in the integral. We use --18 mag in the above calculation, a value which
is often used in the literature \citep[e.g.][]{mad99,gia15,kas15}. 
\citet{gia15} found a sample of very faint AGN candidates with 
$M_{1450}<-18.5$ mag at $z=4\sim6$, and estimated that the faint-end slope of 
the $z=4.75$ AGN luminosity function was roughly --1.81, consistent with our 
result of --1.9 at $z\sim6$. Their faint AGN sample also suggests that we can 
integrate the QLF down to at least $M_{1450}\sim-18$ mag. If we integrate the 
QLF from $M_{1450}=-30$ to --16 mag, the number of ionizing photons increases 
by 17\%\ for the best fits shown in Equation \ref{eq:dplfit}. AGN luminosities 
can be even lower \citep[e.g.][]{ho97,hao05}. 
In the above calculation, we have assumed that the escape fraction of ionizing 
photons is 1. In low-luminosity AGN, however, the escape fraction can be much 
lower \citep[e.g.][]{mic16}. We did not take this into 
account for the luminosity range that we adopted ($M_{1450}<-18$ mag).

The total photon emissivity per unit comoving volume required to ionize the
universe is taken from \citet{mad99}, i.e., 
$\dot{\cal N}_{ion}(z) = 10^{50.48}\,
     \left(\frac{C}{3}\right) \times \left(\frac{1+z}{7}\right)^3 
   {\rm Mpc^{-3}\, s^{-1}},$
where $C$ is the clumping factor of the IGM, and we have assumed that the 
baryon density $\Omega_b h^2=0.022$. The clumping factor $C$ is critical 
as it is closely related to the H recombination rate. Simulations have 
suggested $C\sim2-5$ \citep[e.g.][]{mcq11,fin12,shu12}. In Figure 
\ref{fig:reion}, we show the required photon emissivity (red dashed curves) 
for three representative $C$ values, 1.5, 3, and 5. 
The figure clearly shows the following.
\begin{enumerate}
\item The significance of the quasar contribution to the ionizing background
strongly depends on $\alpha$, $M_{1450}^\ast$, and $C$.
\item If $C=3$, the quasar/AGN population cannot provide enough photons to 
ionize the $z\sim6$ IGM (at $\sim90$\% confidence). We can also rule out
at $\sim68$\% confidence that the quasar/AGN population 
can provide 50\% of the required photons.
\item If $C=3$, the quasar/AGN population can provide enough photons only
if the faint-end slope is significantly steeper than --2 and/or
the characteristic luminosity is very low.
\end{enumerate}

The conclusion that a faint-end slope at the steep end of current 
observations is required for quasars to generate sufficient photons to ionize 
the Universe at $z\sim6$ is generally robust; however, it is worth restating 
some of the key assumptions that went into the above calculations. We assumed 
a constant $M^\ast$, and fixed $k$ and $\beta$ when we calculated the QLF. 
These choices will underestimate the uncertainties in $\alpha$ and $M^{\ast}$. 
We chose a magnitude limit of $M_{1450}=-18$ mag, which extends well below 
current observations. We further assumed an escape fraction of unity for 
ionizing photons, independent of quasar luminosity. Finally, we assumed a 
relatively small range of clumping factors based on current theoretical 
models.

It is generally thought that low-luminosity star-forming galaxies may
provide enough ionizing photons for cosmic reionization, while the 
contribution from quasars/AGN is probably negligible due to their low
spatial density. Figure \ref{fig:reion} shows that we cannot yet fully 
rule out the probability that UV light from AGN is responsible for
ionizing the universe \citep[see also e.g.,][]{gia15,mad15,mit16}; 
however, current observations do not favor this scenario.
Future deep surveys of quasars at $z\ge6$ will provide a definitive answer.

\section{Summary}

In this paper, we have presented the discovery of nine quasars at $z\sim6$ 
identified in the SDSS, including seven quasars in the SDSS main survey, one
quasar in the overlap regions, and one quasar in Stripe 82. One of the quasars 
in the main survey, J1148+0702 at $z=6.339$, is the second highest-redshift 
quasar found in the SDSS. This completes our survey of $z\sim6$ quasars in the 
SDSS footprint. We summarized our final sample of 52 SDSS quasars at $z\sim6$. 
In total, we have found (1) 29 quasars in 11,240 deg$^2$ of the SDSS main 
survey; (2) 17 quasars in 4223 deg$^2$ of the overlap regions (7 of which are 
in common with the main survey sample); (3) 13 quasars in 277 deg$^2$ of 
Stripe 82. The main survey quasars are the most luminous quasars, with 
$z_{\rm AB}\le20$ mag. The overlap region quasars are roughly 0.5 mag fainter,
and the Stripe 82 quasars are 2 mag fainter. The quasars span a wide 
luminosity range of $-29.0\le M_{1450}\le-24.5$ mag, and comprise a 
well-defined quasar sample at $z\sim6$.

Based on the combination of our new quasar sample and two much fainter quasars 
in the literature, we obtained parameters for a double power-law fit to the 
QLF at $z\sim6$ using a maximum likelihood analysis. The bright end of the QLF 
is well constrained, and the slope is steep with $\beta-2.8\pm0.2$. The 
best-fitting results for the faint-end slope $\alpha$ and the characteristic 
magnitude $M_{1450}^{\ast}$ are $\alpha=-1.90_{-0.44}^{+0.58}$ and 
$M^{\ast}=-25.2_{-3.8}^{+1.2}$ mag, and the two quantities are strongly
correlated. The large uncertainties are due to the 
small number of quasars at the faint end.
We calculated the cumulative density of luminous quasars ($M_{1450}\le-26.0$ 
mag, $M_{1450}\le-26.7$ mag, and $M_{1450}\le-25.3$ mag) at $z\sim6$ and 7,
and compared them with those at $z\sim4$ and 5. We found that the cumulative 
density at $z>4$ declines rapidly towards higher redshift. 
We estimated the quasar contribution to the ionizing background
at $z\sim6$ using the derived QLF. 
Assuming an IGM clumping factor $C=3$, the quasar population cannot provide 
enough photons to ionize the $z\sim6$ IGM (at $\sim90$\% confidence).
We found that quasars can provide enough photons only if the faint-end slope
is steeper than --2 and/or if the characteristic luminosity is
very low. A large sample of very faint quasars ($M_{1450}<-23$ mag) is needed
to provide a better constraint on the quasar contribution to cosmic 
reionization.

Many quasars in our sample have been extensively studied in multiple 
wavelength bands from X-ray to radio. More observations are being carried out 
and being planned. For example, we are carrying out deep near-IR spectroscopy 
of $\sim60$ quasars at $z\sim6$ using Gemini GNIRS. When this program 
completes, we will have near-IR (or rest-frame UV) spectra for all the SDSS 
quasars. These spectra will allow us to measure various quasar properties, 
such as spectral index and emission line properties, metallicity in the 
broad-line region, central black-hole mass, and so on. The well-defined sample 
will enable us to derive the black-hole mass function
at $z\sim6$, and further constrain the birth and growth of the earliest
massive black holes. We are also gathering mm/submm observations using
ALMA, IRAM, and JCMT for our SDSS sample \citep[e.g.][]{wang13,wang16b}.
These observations provide rich information on dust emission, star formation,
and dynamical properties of quasar host galaxies, in the context of
galaxy -- black hole co-evolution at the early epoch.
Therefore, this unique SDSS sample will have a legacy value for exploring the
distant universe in the future.

Meanwhile, ongoing and future large-area surveys are finding high-redshift
quasars in large numbers. For example, Pan-STARRS1 has found more than
100 quasars \citep{ban16}, and will improve the constraint on the bright-end
QLF at $z\sim6$. Note that the measurement of the quasar density at the 
brightest end ($M_{1450}<-26$ mag) will not be improved by more than
a factor of two, since the SDSS already covers one fourth of the whole sky.
With near-IR imaging, the VISTA VIKING survey is able to find higher redshift
quasars. It has found 3 quasars at $z>6.5$, and is expecting to find nearly
20 quasars at $6.5<z<7.4$ in the near future \citep{ven13}.
DES, with imaging data slightly deeper than Stripe 82 over $5000$ deg$^2$,
has found its first $z\sim6$ quasar, and has claimed that it would find
50--100 quasars at $z>6$, including 3--10 quasars at $z>7$ \citep{reed15}.
The Subaru SHELLQ survey is producing a large number of very faint quasars 
using deep Hyper Suprime-Cam imaging data \citep{mat16}. These quasars will be 
used to constrain the faint end of the QLF at $z\sim6$. The future LSST survey 
\citep{ive08} will have an
unprecedented power for searches of high-redshift quasars. It will eventually
find thousands of quasars (assuming that there will be enough resources for 
follow-up identification), and fully constrain the $z\sim6$ QLF.
The above surveys are providing a golden opportunity for studying 
high-redshift quasars and the distant universe.

\acknowledgments

We acknowledge support from the Ministry of Science and Technology of China
under grant 2016YFA0400703, and from the National Science Foundation of China 
under grant 11533001. LJ acknowledges support from the Thousand Youth 
Talents Program of China. 
IDM and XF acknowledge support from NSF Grants AST 11-07682 and 15-15115.
YS acknowledges support from an Alfred P. Sloan Research Fellowship.
We thank N. Kashikawa, C. Willott, D.J. Mortlock, B.P. Venemans, and S.J. 
Warren for providing information for Figures \ref{fig:lf2} and \ref{fig:qsoev}.
This research uses data obtained through the Telescope Access Program (TAP), 
which has been funded by the National Astronomical Observatories of China
(NAOC), the Chinese Academy of Sciences (the Strategic Priority Research 
Program ``The Emergence of Cosmological Structures" Grant No. XDB09000000), 
and the Special Fund for Astronomy from the Ministry of Finance. Observations 
obtained with the Hale Telescope at Palomar Observatory were obtained as part 
of an agreement between NAOC and the California Institute of Technology.
Observations reported here were obtained in part at the MMT Observatory,
a joint facility of the University of Arizona and the Smithsonian Institution.
This work is based in part on data obtained as part of the UKIRT Infrared Deep
Sky Survey. 

Funding for the SDSS and SDSS-II has been provided by the Alfred P. Sloan 
Foundation, the Participating Institutions, the National Science Foundation, 
the U.S. Department of Energy, the National Aeronautics and Space 
Administration, the Japanese Monbukagakusho, the Max Planck Society, and the 
Higher Education Funding Council for England. The SDSS Web Site is 
http://www.sdss.org/. 
The SDSS is managed by the Astrophysical Research Consortium for the 
Participating Institutions. The participating institutions are the American 
Museum of Natural History, Astrophysical Institute Potsdam, University of 
Basel, University of Cambridge, Case Western Reserve University, University of 
Chicago, Drexel University, Fermilab, the Institute for Advanced Study, the 
Japan Participation Group, Johns Hopkins University, the Joint Institute for 
Nuclear Astrophysics, the Kavli Institute for Particle Astrophysics and 
Cosmology, the Korean Scientist Group, the Chinese Academy of Sciences 
(LAMOST), Los Alamos National Laboratory, the Max-Planck-Institute for 
Astronomy (MPIA), the Max-Planck-Institute for Astrophysics (MPA), New Mexico 
State University, Ohio State University, University of Pittsburgh, University 
of Portsmouth, Princeton University, the United States Naval Observatory, and 
the University of Washington.

\facilities{Bok (90Prime), MMT (SWIRC, Red Channel Spectrograph), Hale (DBSP),
LBT (MODS), Gemini:North (GNIRS), Magellan:Baade (FIRE), SDSS}

\software{IRAF, IDL}

\appendix

\section{The optical spectra of the 52 SDSS quasars at $z\sim6$}

\begin{figure}
\epsscale{1.2}
\plotone{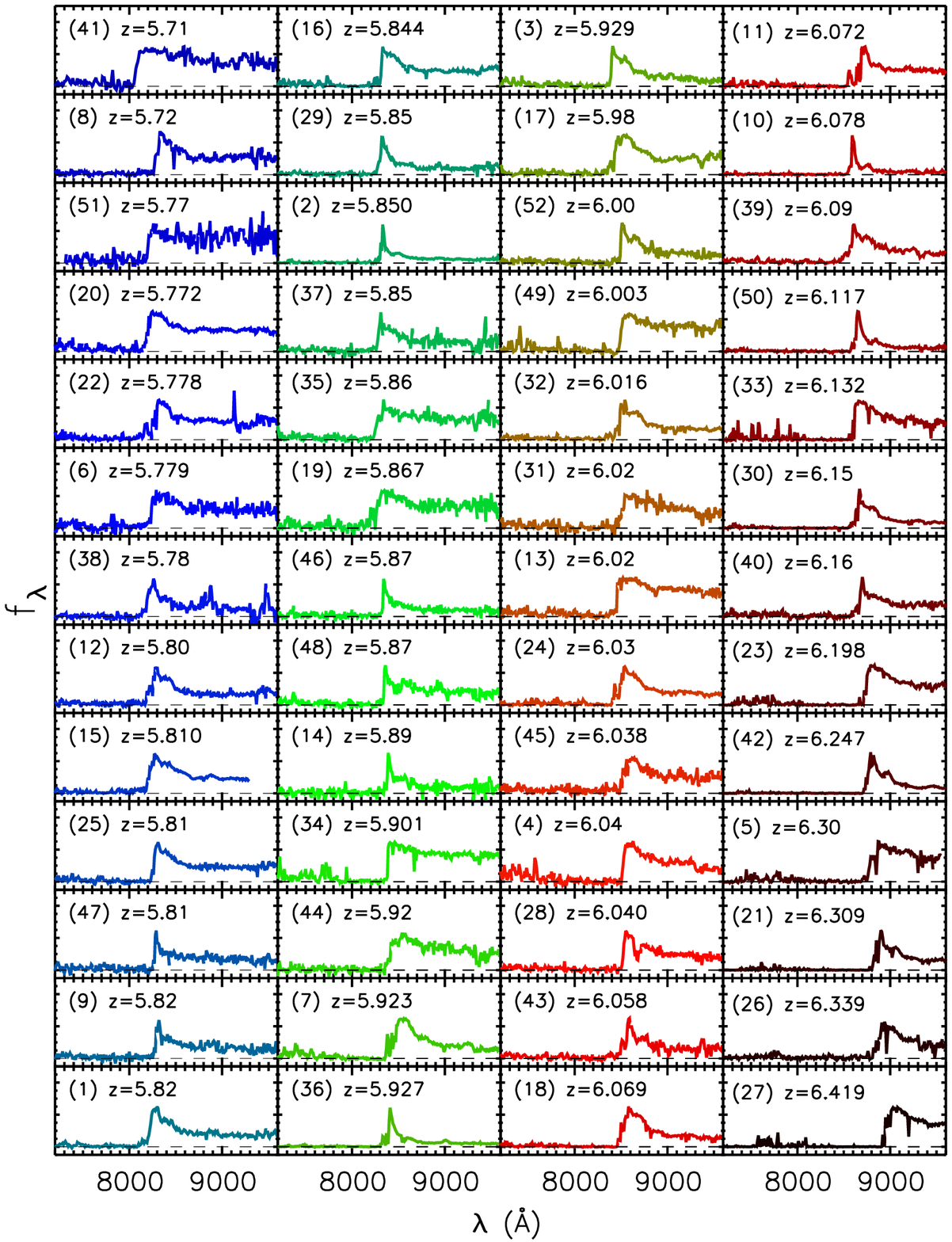}
\caption{Optical spectra of the 52 SDSS quasars at $z\sim6$. 
See the main text for details.
\label{fig:allspec}}
\end{figure}

Figure \ref{fig:allspec} shows the optical spectra of the 52 SDSS quasars at
$z\sim6$, ordered by redshift. The numbers of the quasars correspond to the 
numbers in Column 1 of Table 2. Most of the spectra were taken from the quasar 
discovery papers listed in Column 13 of Table 2. All spectra have been binned 
to 10 \AA\ per pixel. The figure can be downloaded from
\url{http://kiaa.pku.edu.cn/~jiang/SDSS52spectra.eps}, or
\url{http://kiaa.pku.edu.cn/~jiang/SDSS52spectra2.eps}.

\begin{thebibliography}{}
\bibitem[Adelman-McCarthy et al.(2007)]{ade07} Adelman-McCarthy, J.~K.,
   Ag{\"u}eros, M.~A., Allam, S.~S., et al.\ 2007, \apjs, 172, 634
\bibitem[Aihara et al.(2011)]{aih11} Aihara, H., Allende Prieto, C., An, D., 
	et al.\ 2011, \apjs, 193, 29 
\bibitem[Annis et al.(2014)]{ann14} Annis, J., Soares-Santos, M., Strauss,
   M.~A., et al.\ 2014, \apj, 794, 120
\bibitem[Ba{\~n}ados et al.(2014)]{ban14} Ba{\~n}ados, E.,
   Venemans, B.~P., Morganson, E., et al.\ 2014, \aj, 148, 14
\bibitem[Ba{\~n}ados et al.(2015)]{ban15} Ba{\~n}ados, E., Decarli, R., 
	Walter, F., et al.\ 2015, \apjl, 805, L8 
\bibitem[Ba{\~n}ados et al.(2016)]{ban16} Ba{\~n}ados, E., et al.\ 2016,
	\apjs, in press (arXiv:1608.03279)
\bibitem[Barnett et al.(2015)]{bar15} Barnett, R., Warren, S.~J., 
	Banerji, M., et al.\ 2015, \aap, 575, A31 
\bibitem[Becker et al.(2001)]{bec01} Becker, R.~H., Fan, X., White, R.~L., 
	et al.\ 2001, \aj, 122, 2850 
\bibitem[Becker et al.(2011)]{bec11} Becker, G.~D., Sargent, W.~L.~W., 
	Rauch, M., \& Calverley, A.~P.\ 2011, \apj, 735, 93 
\bibitem[Becker et al.(2015)]{bec15} Becker, G.~D., Bolton, J.~S., 
	Madau, P., et al.\ 2015, \mnras, 447, 3402 
\bibitem[Blain et al.(2013)]{bla13} Blain, A.~W., 
	Assef, R., Stern, D., et al.\ 2013, \apj, 778, 113 
\bibitem[Bolton et al.(2011)]{bol11} Bolton, J.~S., Haehnelt, M.~G., 
	Warren, S.~J., et al.\ 2011, \mnras, 416, L70 
\bibitem[Brown et al.(2008)]{swirc} Brown, W.~R., McLeod, B.~A., Geary, J.~C., \& Bowsher, E.~C.\ 2008, \procspie, 7014, 70142P 
\bibitem[Calura et al.(2014)]{cal14} Calura, F., Gilli, R., Vignali, C., 
	et al.\ 2014, \mnras, 438, 2765 
\bibitem[Carilli et al.(2010)]{car10} Carilli, C.~L., Wang, R., Fan, X.,
   et al.\ 2010, \apj, 714, 834
\bibitem[Carilli \& Walter(2013)]{car13} Carilli, C.~L., \& Walter, F.\ 
	2013, \araa, 51, 105 
\bibitem[Carnall et al.(2015)]{car15} Carnall, A.~C., Shanks, T., Chehade, B., 
	et al.\ 2015, \mnras, 451, L16 
\bibitem[Croom et al.(2009)]{cro09} Croom, S.~M., Richards, G.~T., Shanks, T., 
	et al.\ 2009, \mnras, 399, 1755 
\bibitem[Cool et al.(2006)]{cool06} Cool, R.~J., Kochanek, C.~S.,
   Eisenstein, D.~J., et al.\ 2006, \aj, 132, 823
\bibitem[De Rosa et al.(2011)]{derosa11} De Rosa, G., Decarli, R.,
   Walter, F., et al.\ 2011, \apj, 739, 56
\bibitem[De Rosa et al.(2014)]{derosa14} De Rosa, G., Venemans, B.~P., 
	Decarli, R., et al.\ 2014, \apj, 790, 145 
\bibitem[DiPompeo et al.(2014)]{dip14} DiPompeo, M.~A., Myers, A.~D., 
	Brotherton, M.~S., Runnoe, J.~C., \& Green, R.~F.\ 2014, \apj, 787, 73 
\bibitem[Fan(1999)]{fan99} Fan, X.\ 1999, \aj, 117, 2528 
\bibitem[Fan et al.(2000)]{fan00} Fan, X., White, R.~L.,
   Davis, M., et al.\ 2000, \aj, 120, 1167
\bibitem[Fan et al.(2001a)]{fan01a} Fan, X., Narayanan, V.~K.,
   Lupton, R.~H., et al.\ 2001, \aj, 122, 2833
\bibitem[Fan et al.(2001b)]{fan01b} Fan, X., Strauss, M.~A., Schneider, D.~P., 
   et al.\ 2001, \aj, 121, 54     
\bibitem[Fan et al.(2003)]{fan03} Fan, X., Strauss, M.~A.,
   Schneider, D.~P., et al.\ 2003, \aj, 125, 1649
\bibitem[Fan et al.(2004)]{fan04} Fan, X., Hennawi, J.~F.,
   Richards, G.~T., et al.\ 2004, \aj, 128, 515
\bibitem[Fan et al.(2006a)]{fan06a} Fan, X., Strauss, M.~A.,
   Richards, G.~T., et al.\ 2006, \aj, 131, 1203
\bibitem[Fan et al.(2006b)]{fan06b} Fan, X., Strauss, M.~A.,
   Becker, R.~H., et al.\ 2006, \aj, 132, 117
\bibitem[Fasano \& Franceschini(1987)]{fas87} Fasano, G., \& Franceschini, 
	A.\ 1987, \mnras, 225, 155 
\bibitem[Finlator et al.(2012)]{fin12} Finlator, K., Oh, S.~P., 
	{\"O}zel, F., \& Dav{\'e}, R.\ 2012, \mnras, 427, 2464 
\bibitem[Fiore et al.(2012)]{fio12} Fiore, F., Puccetti, S., 
	Grazian, A., et al.\ 2012, \aap, 537, A16 
\bibitem[Fliri \& Trujillo(2016)]{fli16} Fliri, J., \& Trujillo, I.\ 2016, 
	\mnras, 456, 1359 
\bibitem[Gallerani et al.(2010)]{gal10} Gallerani, S., Maiolino, R., 
	Juarez, Y., et al.\ 2010, \aap, 523, A85 
\bibitem[G{\'o}rski et al.(2005)]{gor05} G{\'o}rski, K.~M., Hivon, E., 
	Banday, A.~J., et al.\ 2005, \apj, 622, 759 
\bibitem[Goto(2006)]{goto06} Goto, T.\ 2006, \mnras, 371, 769
\bibitem[Fukugita et al.(1996)]{fuk96} Fukugita, M., Ichikawa, T., Gunn,
   J.~E., et al.\ 1996, \aj, 111, 1748
\bibitem[Giallongo et al.(2015)]{gia15} Giallongo, E., Grazian, A., Fiore, 
	F., et al.\ 2015, \aap, 578, A83 
\bibitem[Glikman et al.(2011)]{gli11} Glikman, E., 
	Djorgovski, S.~G., Stern, D., et al.\ 2011, \apjl, 728, L26 
\bibitem[Gunn et al.(1998)]{gun98} Gunn, J.~E., Carr, M., Rockosi, C.,
   et al.\ 1998, \aj, 116, 3040
\bibitem[Gunn et al.(2006)]{gun06} Gunn, J.~E., Siegmund, W.~A., Mannery,
   E.~J., et al.\ 2006, \aj, 131, 2332
\bibitem[Hao et al.(2005)]{hao05} Hao, L., Strauss, M.~A., Fan, X., 
	et al.\ 2005, \aj, 129, 1795 
\bibitem[Ho et al.(1997)]{ho97} Ho, L.~C., Filippenko, A.~V., \& Sargent, 
	W.~L.~W.\ 1997, \apjs, 112, 315 
\bibitem[Hogg et al.(2001)]{hog01} Hogg, D.~W., Finkbeiner,
	D.~P., Schlegel, D.~J., \& Gunn, J.~E.\ 2001, \aj, 122, 2129
\bibitem[Ikeda et al.(2011)]{ike11} Ikeda, H., Nagao, T., Matsuoka, K., 
	et al.\ 2011, \apjl, 728, L25 
\bibitem[Ikeda et al.(2012)]{ike12} Ikeda, H., Nagao, T., Matsuoka, K., 
	et al.\ 2012, \apj, 756, 160 
\bibitem[Ivezi{\'c} et al.(2004)]{ive04} Ivezi{\'c}, {\v Z}., Lupton, 
	R.~H., Schlegel, D., et al.\ 2004, Astronomische Nachrichten, 325, 583
\bibitem[Ivezic et al.(2008)]{ive08} Ivezic, Z., Axelrod, T., 
	Brandt, W.~N., et al.\ 2008, Serbian Astronomical Journal, 176, 1 
\bibitem[Jiang et al.(2006)]{jiang06} Jiang, L., Fan, X., Hines, D.~C., 
	et al.\ 2006, \aj, 132, 2127 
\bibitem[Jiang et al.(2007)]{jiang07} Jiang, L., Fan, X., Vestergaard, M.,
   et al.\ 2007, \aj, 134, 1150
\bibitem[Jiang et al.(2008)]{jiang08} Jiang, L., Fan, X., Annis, J.,
   et al.\ 2008, \aj, 135, 1057
\bibitem[Jiang et al.(2009)]{jiang09} Jiang, L., Fan, X., Bian, F.,
   et al.\ 2009, \aj, 138, 305
\bibitem[Jiang et al.(2010)]{jiang10} Jiang, L., Fan, X., Brandt, W.~N.,
   et al.\ 2010, \nat, 464, 380
\bibitem[Jiang et al.(2014)]{jiang14} Jiang, L., Fan, X., Bian, F.,
   et al.\ 2014, \apjs, 213, 12
\bibitem[Jiang et al.(2015)]{jiang15} Jiang, L., McGreer, I.~D.,
   Fan, X., et al.\ 2015, \aj, 149, 188
\bibitem[Juarez et al.(2009)]{jua09} Juarez, Y., Maiolino, R., 
	Mujica, R., et al.\ 2009, \aap, 494, L25 
\bibitem[Jun et al.(2015)]{jun15} Jun, H.~D., Im, M., Lee, H.~M., 
	et al.\ 2015, \apj, 806, 109 
\bibitem[Lusso et al.(2015)]{lus15} Lusso, E., Worseck, G., 
	Hennawi, J.~F., et al.\ 2015, \mnras, 449, 4204 
\bibitem[Kaiser et al.(2010)]{kai10} Kaiser, N., Burgett, W., 
	Chambers, K., et al.\ 2010, \procspie, 7733, 77330E 
\bibitem[Kashikawa et al.(2015)]{kas15} Kashikawa, N., Ishizaki, Y., 
	Willott, C.~J., et al.\ 2015, \apj, 798, 28 
\bibitem[Kim et al.(2015)]{kim15} Kim, Y., Im, M., Jeon, Y., 
	et al.\ 2015, \apjl, 813, L35 
\bibitem[Kurk et al.(2007)]{kurk07} Kurk, J.~D., Walter, F., Fan, X.,
   et al.\ 2007, \apj, 669, 32
\bibitem[Kurk et al.(2009)]{kurk09} Kurk, J.~D., Walter, F., Fan, X., 
	et al.\ 2009, \apj, 702, 833 
\bibitem[Leipski et al.(2014)]{lei14} Leipski, C., Meisenheimer, K., 
	Walter, F., et al.\ 2014, \apj, 785, 154 
\bibitem[Lupton et al.(2001)]{lup01} Lupton, R., Gunn, J.~E., Ivezi{\'c}, Z., 
	Knapp, G.~R., \& Kent, S.\ 2001, Astronomical Data Analysis Software and 
	Systems X, 238, 269 
\bibitem[Lyu et al.(2016)]{lyu16} Lyu, J., Rieke, G.~H., \& Alberts, S.\ 
	2016, \apj, 816, 85 
\bibitem[Marshall et al.(1983)]{mar83} Marshall, H. L., Avni, Y., Tananbaum, 
	H., \& Zamorani, G. 1983, \apj, 269, 35
\bibitem[Madau et al.(1999)]{mad99} Madau, P., Haardt, F., \& Rees, M.~J.\
  1999, \apj, 514, 648
\bibitem[Madau \& Haardt(2015)]{mad15} Madau, P., \& Haardt, F.\ 2015, 
	\apjl, 813, L8 
\bibitem[Matsuoka et al.(2016)]{mat16} Matsuoka, Y., Onoue, M.,
   Kashikawa, N., et al.\ 2016, \apj, 828, 26
\bibitem[McGreer et al.(2006)]{mcg06} McGreer, I.~D., Becker, R.~H.,
   Helfand, D.~J., \& White, R.~L.\ 2006, \apj, 652, 157
\bibitem[McGreer et al.(2013)]{mcg13} McGreer, I.~D., Jiang, L., Fan, X., 
	et al.\ 2013, \apj, 768, 105 
\bibitem[McGreer et al.(2015)]{mcg15} McGreer, I.~D., Mesinger, A., 
	\& D'Odorico, V.\ 2015, \mnras, 447, 499 
\bibitem[McQuinn et al.(2011)]{mcq11} McQuinn, M., Oh, S.~P., \& 
	Faucher-Gigu{\`e}re, C.-A.\ 2011, \apj, 743, 82 
\bibitem[Micheva et al.(2016)]{mic16} Micheva, G., Iwata, I., \& 
	Inoue, A.~K.\ 2016, \mnras, in press (arXiv:1604.00102)
\bibitem[Mitra et al.(2016)]{mit16} Mitra, S., Choudhury, T.~R., \& 
	Ferrara, A.\ 2016, \mnras, summited (arXiv:1606.02719) 
\bibitem[Morganson et al.(2012)]{morg12} Morganson, E., De Rosa, G.,
   Decarli, R., et al.\ 2012, \aj, 143, 142
\bibitem[Mortlock et al.(2009)]{mort09} Mortlock, D.~J., Patel, M.,
   Warren, S.~J., et al.\ 2009, \aap, 505, 97
\bibitem[Mortlock et al.(2011)]{mort11} Mortlock, D.~J., Warren, S.~J.,
   Venemans, B.~P., et al.\ 2011, \nat, 474, 616
\bibitem[Mortlock(2015)]{mort15} Mortlock, D.~J.\ 2015, arXiv:1511.01107 
\bibitem[Omont et al.(2013)]{omo13} Omont, A., Willott, C.~J., Beelen, A., 
	et al.\ 2013, \aap, 552, A43 
\bibitem[Padmanabhan et al.(2008)]{pad08} Padmanabhan, N., Schlegel, D.~J.,
   Finkbeiner, D.~P., et al.\ 2008, \apj, 674, 1217
\bibitem[Peacock(1983)]{pea83} Peacock, J.~A.\ 1983, \mnras, 202, 615 
\bibitem[Pier et al.(2003)]{pier03} Pier, J.~R., Munn, J.~A., Hindsley, 
	R.~B., et al.\ 2003, \aj, 125, 1559 
\bibitem[Plotkin et al.(2015)]{plo15} Plotkin, R.~M., Shemmer, O., 
	Trakhtenbrot, B., et al.\ 2015, \apj, 805, 123 
\bibitem[Reed et al.(2015)]{reed15} Reed, S.~L., McMahon, R.~G., Banerji, M., 
	et al.\ 2015, \mnras, 454, 3952 
\bibitem[Richards et al.(2006)]{ric06} Richards, G.~T., Strauss, M.~A., 
	Fan, X., et al.\ 2006, \aj, 131, 2766 
\bibitem[Ross et al.(2012)]{ross12} Ross, N.~P., Myers, A.~D., Sheldon, 
	E.~S., et al.\ 2012, \apjs, 199, 3 
\bibitem[Ross et al.(2013)]{ross13} Ross, N.~P., McGreer, I.~D., White, M., 
	et al.\ 2013, \apj, 773, 14 
\bibitem[Schmidt et al.(1989)]{mmtred} Schmidt, G.~D., Weymann, R.~J., \& Foltz, C.~B.\ 1989, \pasp, 101, 713 
\bibitem[Shen et al.(2007)]{shen07} Shen, Y., Strauss, M.~A., Oguri, M.,
   et al.\ 2007, \aj, 133, 2222
\bibitem[Shen \& Liu(2012)]{shen12} Shen, Y., \& Liu, X.\ 2012, \apj, 753, 125 
\bibitem[Shull et al.(2012)]{shu12} Shull, J.~M., Harness, A., Trenti, M., 
	\& Smith, B.~D.\ 2012, \apj, 747, 100 
\bibitem[Simcoe et al.(2011)]{sim11} Simcoe, R.~A., Cooksey, K.~L., 
	Matejek, M., et al.\ 2011, \apj, 743, 21 
\bibitem[Skrutskie et al.(2006)]{skr06} Skrutskie, M.~F., Cutri, R.~M.,
   Stiening, R., et al.\ 2006, \aj, 131, 1163
\bibitem[Smith et al.(2002)]{smi02} Smith, J.~A., Tucker, D.~L., 
	Kent, S., et al.\ 2002, \aj, 123, 2121 
\bibitem[Strauss et al.(1999)]{str99} Strauss, M.~A., Fan, X., 
	Gunn, J.~E., et al.\ 1999, \apjl, 522, L61 
\bibitem[Tucker et al.(2006)]{tuc06} Tucker, D.~L., Kent, S., Richmond, M.~W., 
	et al.\ 2006, Astronomische Nachrichten, 327, 821 
\bibitem[Venemans et al.(2007)]{ven07} Venemans, B.~P., McMahon, R.~G.,
   Warren, S.~J., et al.\ 2007, \mnras, 376, L76
\bibitem[Venemans et al.(2013)]{ven13} Venemans, B.~P., Findlay, J.~R., 
	Sutherland, W.~J., et al.\ 2013, \apj, 779, 24 
\bibitem[Venemans et al.(2015)]{ven15} Venemans, B.~P., Ba{\~n}ados, E.,
   Decarli, R., et al.\ 2015, \apjl, 801, L11
\bibitem[Walter et al.(2009)]{wal09} Walter, F., Riechers,
   D., Cox, P., et al.\ 2009, \nat, 457, 699
   et al.\ 2013, \apj, 773, 44
\bibitem[F. Wang et al.(2016)]{wang16a} Wang, F., Wu, X.-B., Fan, X.,
   et al.\ 2016, \apj, 819, 24
\bibitem[Wang et al.(2008)]{wang08} Wang, R., Carilli, C.~L., 
	Wagg, J., et al.\ 2008, \apj, 687, 848-858 
\bibitem[Wang et al.(2011)]{wang11} Wang, R., Wagg, J., Carilli, C.~L.,
   et al.\ 2011, \aj, 142, 101
\bibitem[Wang et al.(2013)]{wang13} Wang, R., Wagg, J., Carilli, C.~L.,
   et al.\ 2013, \apj, 773, 44
\bibitem[R. Wang et al.(2016)]{wang16b} Wang, R., Wu, X.-B., Neri, R.,
	et al.\ 2016, \apjl, in press (arXiv:1606.09634)
\bibitem[Warren et al.(2007)]{war07} Warren, S.~J., Hambly, N.~C.,
   Dye, S., et al.\ 2007, \mnras, 375, 213
\bibitem[White et al.(2003)]{whi03} White, R.~L., Becker, R.~H., Fan, X.,
   \& Strauss, M.~A.\ 2003, \aj, 126, 1
\bibitem[Willott et al.(2007)]{wil07} Willott, C.~J., Delorme, P.,
   Omont, A., et al.\ 2007, \aj, 134, 2435
\bibitem[Willott et al.(2009)]{wil09} Willott, C.~J., Delorme, P.,
   Reyl{\'e}, C., et al.\ 2009, \aj, 137, 3541
\bibitem[Willott et al.(2010a)]{wil10a} Willott, C.~J., Albert, L., Arzoumanian, D., et al.\ 2010a, \aj, 140, 546 
\bibitem[Willott et al.(2010b)]{wil10} Willott, C.~J., Delorme, P.,
   Reyl{\'e}, C., et al.\ 2010b, \aj, 139, 906
\bibitem[Wright et al.(2010)]{wri10} Wright, E.~L., 
	Eisenhardt, P.~R.~M., Mainzer, A.~K., et al.\ 2010, \aj, 140, 1868-1881 
\bibitem[Wu et al.(2015)]{wu15} Wu, X.-B., Wang, F., Fan, X.,
   et al.\ 2015, \nat, 518, 512
\bibitem[Wyithe \& Bolton(2011)]{wyi11} Wyithe, J.~S.~B., \& Bolton, J.~S.\ 
	2011, \mnras, 412, 1926 
\bibitem[Yan et al.(2013)]{yan13} Yan, L., Donoso, E., Tsai, C.-W., et al.\ 
	2013, \aj, 145, 55 
\bibitem[Yang et al.(2016)]{yang16} Yang, J., Wang, F., Wu, X., et al.\ 2016,
	\apj, in press (arXiv:1607.04415)
\bibitem[York et al.(2000)]{york00} York, D.~G., Adelman, J.,
   Anderson, J.~E., Jr., et al.\ 2000, \aj, 120, 1579
\bibitem[Zeimann et al.(2011)]{zei11} Zeimann, G.~R., White, R.~L., 
	Becker, R.~H., et al.\ 2011, \apj, 736, 57
\end{thebibliography}
\end{document}